\documentclass[twocolumn]{aastex63}

\pdfoutput=1

\usepackage{amsfonts}
\usepackage{amssymb}
\usepackage{gensymb}
\usepackage{upgreek}
\usepackage{graphicx}
\usepackage{enumerate}
\usepackage{verbatim}
\usepackage[normalem]{ulem}

\begin{document}

\title{Measuring Dust Attenuation Curves of SINGS/KINGFISH Galaxies Using {\it Swift}/UVOT Photometry}

\author[0000-0002-8606-0797]{Alexander Belles}
\affiliation{Department of Astronomy \& Astrophysics, The Pennsylvania
State University, University Park, PA 16802}

\author[0000-0001-9462-5543]{Marjorie Decleir}
\affiliation{Space Telescope Science Institute, 3700 San Martin Drive, Baltimore, MD 21218}

\author[0000-0003-4381-5245]{William P. Bowman}
\affiliation{Department of Astronomy \& Astrophysics, The Pennsylvania
State University, University Park, PA 16802}
\affiliation{Institute for Gravitation and the Cosmos, The Pennsylvania State University, University Park, PA 16802}
\affiliation{Department of Astronomy, Yale University, New Haven, CT 06520}

\author[0000-0001-8918-1597]{Lea M.Z. Hagen}
\affiliation{Space Telescope Science Institute, 3700 San Martin Drive, Baltimore, MD 21218}

\author[0000-0001-6842-2371]{Caryl Gronwall}
\affiliation{Department of Astronomy \& Astrophysics, The Pennsylvania
State University, University Park, PA 16802}
\affiliation{Institute for Gravitation and the Cosmos, The Pennsylvania State University, University Park, PA 16802}

\author[0000-0003-1817-3009]{Michael H. Siegel}
\affiliation{Department of Astronomy \& Astrophysics, The Pennsylvania
State University, University Park, PA 16802}


\correspondingauthor{A. Belles} 
\email{aub1461@psu.edu}

\begin{abstract}
    We present {\it Swift}/Ultraviolet Optical Telescope (UVOT) integrated light photometry of the {\it Spitzer} Infrared Nearby Galaxies Survey (SINGS) and the Key Insights on Nearby Galaxies: A Far-Infrared Survey with {\it Herschel} (KINGFISH) samples of nearby galaxies.  Combining the {\it Swift}/UVOT data with archival photometry, we investigate a variety of dust attenuation curves derived using \texttt{MCSED}, a flexible spectral energy distribution fitting code. We fit the panchromatic data using three different star formation history (SFH) parameterizations: a decaying exponential, a double power law, and a piecewise function with breaks at physically motivated ages. We find that the average attenuation law of the sample changes slightly based on the SFH assumed. Specifically, the exponential SFH leads to the shallowest attenuation curves. Using simulated data, we also find the exponential SFH fails to outperform the more complex SFHs. Finally, we find a systematic offset in the derived bump strength between SED fits with and without UVOT data, where the inclusion of UVOT data leads to smaller bump strengths, highlighting the importance of the UVOT data. This discrepancy is not seen in fits to mock photometry. Understanding dust attenuation in the local universe is key to understanding high redshift objects where rest-frame far-infrared data is unavailable. 
\end{abstract}


\section{Introduction} \label{sec:intro}

Dust is ubiquitous in the interstellar medium (ISM) but comprises only a small fraction of the mass of a galaxy \citep{Galliano_dustreview}. However, dust plays an oversized role in affecting the light from a galaxy; light emitted in a galaxy will be absorbed by interstellar dust and re-radiated in the infrared (IR). The wavelength dependence of the extinguishing of stellar light, called the dust extinction law, has been studied within the Milky Way (MW) and an average extinction law is typically used \citep[e.g.][]{ccm89, ODonnell_94, F99, Draine_2003, 2004ApJ...616..912V, 2013ApJ...771...68P, 2019ApJ...886..108F,  2020ApJ...891...67M, 
2021ApJ...916...33G, 2022ApJ...930...15D}. However, these works also show variations in the extinction law for different sight lines. Using an average extinction law can introduce systematic uncertainty to the ultraviolet, where the amount of extinction increases rapidly with decreasing wavelength. Since young, UV-bright stars outshine older stellar populations, it is imperative to understand the effects of dust. 

Our knowledge of how dust in other galaxies affects their own stellar emission is even more uncertain due to the effects of star-dust geometry. For stars in the MW, the geometry is a screen of dust between the source and the observer and the radiative transfer is simple; the intensity of starlight decreases exponentially based on the optical depth. The amount, composition, and particle size distribution of the dust solely determine the extinction law. For extended objects like other galaxies, the star-dust geometry is more complex. Light can be scattered into the line of sight and sources are affected by differing column depths of dust. This is referred to as attenuation and the radiative transfer effects are much more complicated \citep{2001PASP..113.1449C, 2013ARA&A..51...63S}. Due to the complexity, it is typical to assume an effective screen of dust that affects all light equally. In certain cases, additional attenuation is added to young stars to account for their birth cloud \citep{CF00}.

The effects of dust in nearby galaxies have been most comprehensively studied in the Small and Large Magellanic Clouds (SMC/LMC) and starburst galaxies \citep{G03, C94, C00}. Due to the proximity of the Magellanic Clouds, individual stars can be resolved and extinction curves can be measured by comparing the extinguished stellar spectrum to a known or theoretical unreddened spectrum of the same spectral type (known as the pair method). The Starburst law from \citet{C00} was derived through similar means using UV and optical spectra and the Balmer decrement, which measures the deviation of the ratio $I(H_{\alpha})/I(H_{\beta})$ from its theoretical value of 2.86, to quantify the effects of dust. 

\begin{figure}
    \centering
    \includegraphics[width=0.5\textwidth]{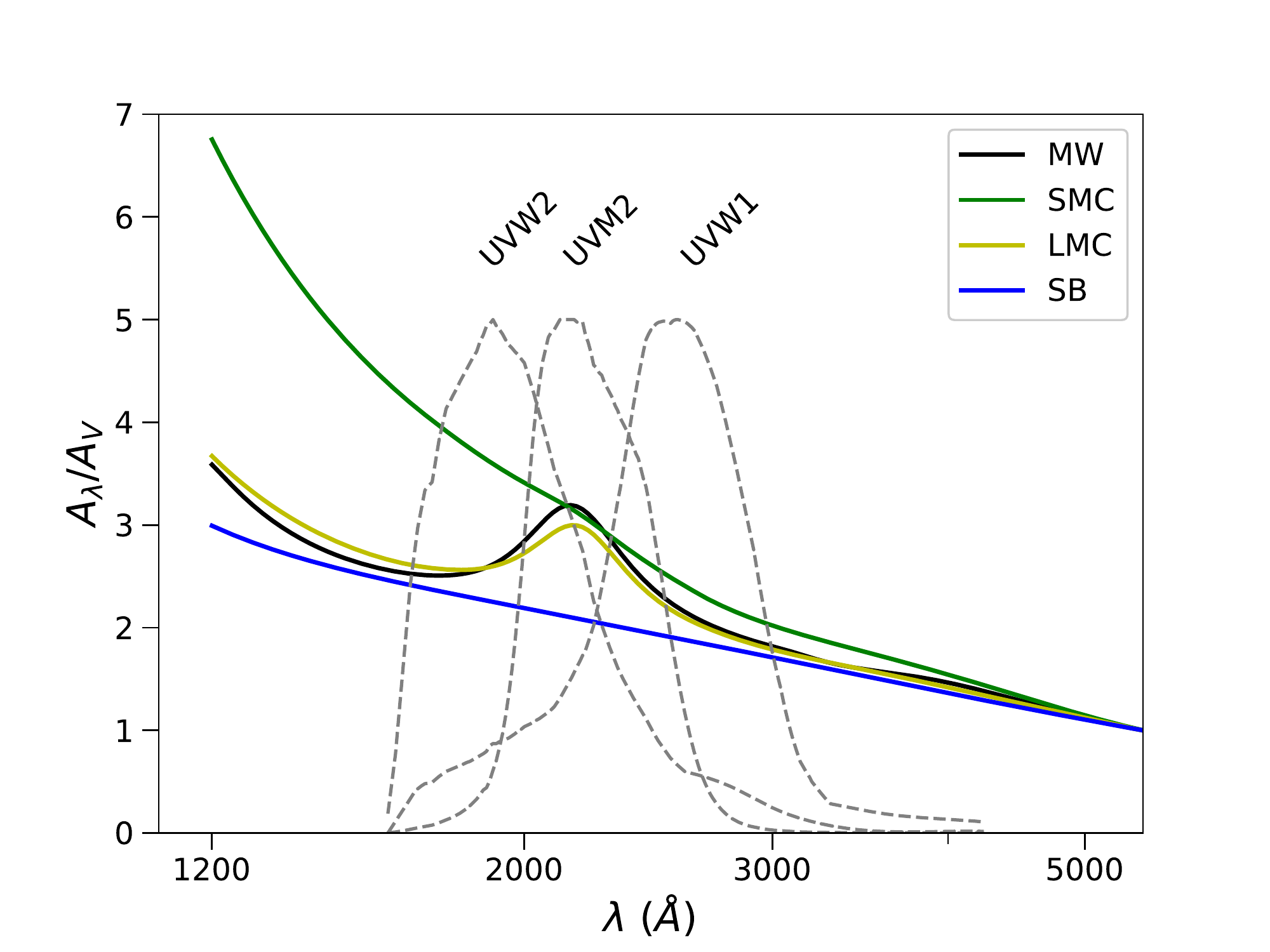}
    \caption{Measured dust laws and the transmission curves of the three {\it Swift} NUV filters. SMC/LMC (green and yellow respectively) from \citet{G03}, MW (black) from \citet{ccm89}, Starburst (SB; blue) from \citet{C00}. Filter transmission curves of the 3 NUV {\it Swift} UVOT filters taken from the SVO Filter Profile Service \citep{2020sea..confE.182R} are overlaid on the measured dust laws (gray dashed lines). The UVM2 filter covers the 2175 \AA\ bump and the UVW1 and UVW2 filters are on either side, allowing us to better measure the UV SED and constrain the UV attenuation law. Note: the MW extinction law and other galaxy attenuation laws are not directly comparable.}
    \label{fig:dustlaws}
\end{figure}

The Starburst law differed greatly from other measured dust laws. Figure~\ref{fig:dustlaws} compares the measured dust law for the MW, Starbursts, and the Magellanic clouds \citep{ccm89, C94, C00, G03}\footnote{Comparing dust extinction to attenuation in other galaxies is not a one-to-one comparison but is still illustrative.}. The biggest differences between the different dust laws are the UV slope and existence of excess extinction at 2175 \AA, a feature called the UV bump which is thought to be from some carbonaceous species, abundant throughout the ISM and first detected in other galaxies by \citet{2010ApJ...718..184C} \citep{1965ApJ...142.1683S, 2011piim.book.....D, 2020ARA&A..58..529S}. Due to the variable slopes and 2175 \AA\ bump strengths seen, a modified Starburst law, introduced in \citet{Noll_09} and further modified in \citet{Kriek+Conroy}, allows for a power law deviation $\delta$ anchored at 5500 \AA\ and a Lorentzian-like Drude profile bump at 2175 \AA\ with a fixed width and variable strength $E_b$. The stark difference at the bluest wavelengths raises questions about how accurate galaxy properties can be measured at the bluest wavelengths due to the systematic uncertainty introduced by dust.

A modern method to deduce attenuation laws from observations of other galaxies is spectral energy distribution (SED) fitting, where broadband observed flux measurements are fit using theoretical model spectra of stellar populations including the effects of attenuation and thermal dust emission \citep[see the reviews,][]{Conroy_SED, 2020IAUS..341...26B}. This multi-faceted problem requires many assumptions to model all possible emission and absorption sources, as well as assuming forms for the star formation history and initial mass function. In addition to the necessary assumptions, SED fitting is often performed using the total integrated light from galaxies, removing any information arising from local structures within a galaxy. Many SED fitting codes exist based on varying assumptions such as CIGALE, GRASIL, MAGPHYS, BAGPIPES, and Prospector \citep{Noll_09, 2019A&A...622A.103B, GRASIL, MAGPHYS, 2018MNRAS.480.4379C, 2017ApJ...837..170L, 2021ApJS..254...22J}.

Theoretical work has shown that the attenuation law can vary from galaxy to galaxy due to the star-dust geometry. \citet{2018ApJ...869...70N} modeled the variety of attenuation laws that could be observed at different redshifts, assuming a single, fixed dust extinction model. They found dramatic variations in the inferred attenuation laws due to the radiative transfer effects arising from different dust-star geometries. \citet{2020MNRAS.491.3937T} provide a model of dust attenuation from a cosmological simulation based on optical depth and dust surface density. Similarly, \citet{2022ApJ...931...14L} showed that varying the fraction of the light unobscured by dust, mimicking complex geometries, can improve the ability to recover the known attenuation law from simulated galaxy SEDs. 

Many observational SED fitting studies have found relationships between attenuation curve parameters and galaxy properties such as stellar mass, inclination, star formation rate. \citet{Kriek+Conroy} stacked photometry of galaxies of the same spectral type over a range of redshifts to produce well-sampled composite SEDs. Their SED fitting elucidated a relationship between the slope of the attenuation law and the bump strength, finding that galaxies with shallow attenuation laws have minimal dust bumps and later explained in the theoretical work of \citet{2018ApJ...869...70N}. Works such as  \citet{2011MNRAS.417.1760W} and \citet{Salim_2018} derived the dust attenuation laws for a large sample of nearby galaxies and found relationships between the slope of the attenuation law and the galaxies' optical opacity. Other recent studies have explored the effect galaxy inclination has on the attenuation law and how it relates to the star-dust geometry \citep{2017ApJ...851...90B, 2021ApJ...923...26D, 2021ApJ...922L..32Z, 2022ApJ...932...54N}. However, there is not a clear consensus in the literature as the scatter in these relationships are large \citep[e.g.][]{decleir_thesis, 2020ARA&A..58..529S, 2022ApJ...932...54N}. Trends with other galaxy properties like metallicity and PAH features have also been studied \citep{2020ApJ...899..117S, 2022MNRAS.514.1886S}. 

Despite the large variety of attenuation curves measured, the starburst law is commonly assumed for extragalactic sources. This can be problematic in the UV where assuming the grey starburst law can lead to underestimates in galaxy properties like star formation rate (SFR). In the context of high redshift galaxies, other means to constrain the star formation rate can be more difficult to obtain, making dust corrections all the more important. Some studies have derived attenuation curves at moderate redshifts, allowing study of its time evolution \citep{2018MNRAS.475.2363T, 2019MNRAS.488.2301T, 2020ApJ...888..108B, 2020ApJ...899..117S, 2022MNRAS.514.1886S}. However, it is difficult to statistically study the highest redshift objects. A poor UV attenuation correction can greatly affect precision science at high redshift.

To better constrain the UV attenuation curve, several studies have used the Ultraviolet Optical Telescope (UVOT) on the Neil Gehrels {\it Swift} Observatory to provide additional UV coverage \citep{swift_ref, UVOT_ref}. For example, \citet{hoversten} and \citet{2014MNRAS.440..150H, 2015MNRAS.452.1412H} looked at dust properties in M81 and M82 respectively; \citet{Hagen_SMC} studied the variation in the dust law across the SMC; and more recently, \citet{Decleir2019}, \citet{decleir_thesis}, \citet{mallory}, and \citet{2022arXiv221201918Z} have used UVOT to study attenuation in nearby galaxies on a spatially resolved scale. UVOT has also been used for studying the extinction law within the MW \citep{2021MNRAS.505..283F}.

In this work, we further investigate the diversity of UV attenuation curves in nearby galaxies by adding {\it Swift}/UVOT to the SINGS/KINGFISH sample \citep{SINGS_2003PASP, KINGFISH_2011PASP..123.1347K}. Given the wavelength range of UVOT, the additional coverage will help constrain both the slope of the attenuation curve and the 2175 \AA\ dust bump (see Figure~\ref{fig:dustlaws}). We derive UV attenuation curves via SED fitting, specifically using MCSED, a flexible SED fitting code \citep{2020ApJ...899....7B}.

Using the \citet{Noll_09} and \citet{Kriek+Conroy} parameterization of the attenuation law, we look to understand how the assumed form of the star formation history (SFH) affects the derived attenuation law. The SFH is not constrained by photometry but imposes strong priors on the ratio of young to old stars. Since young stars are luminous in the UV, the derived attenuation law will change based on the number of young stars. Previous works have studied the effect of parametric and nonparametric forms of the SFH on properties like stellar mass and star formation rate, with nonparametric SFHs performing best due to their flexibility \citep{2019ApJ...873...44C, 2019ApJ...876....3L, lower20, 2022ApJ...935..146S}. 

The paper is organized as follows. In Section~\ref{sec:data}, we discuss our sample and the data used. In Section~\ref{sec:methods}, we describe our data reduction and photometry measurement for {\it Swift}/UVOT. Section~\ref{sec:SED_fitting} discusses our SED modeling and the various assumptions made. In Sections~\ref{sec:results} and~\ref{sec:discussion}, we present and discuss our results, and then summarize our conclusions in Section~\ref{sec:conclusion}. 

In this paper, we assume a \citet{kroupa} Initial Mass Function (IMF) and a $\Lambda$CDM cosmology with $\Omega_m = 0.3$, $\Omega_\Lambda = 0.7$, and $H_0 = 70$ km s$^{-1}$ Mpc$^{-1}$.

\section{Sample \& Data} \label{sec:data}

\begin{figure*}[!htbp]
    \centering
    \includegraphics[width=\textwidth]{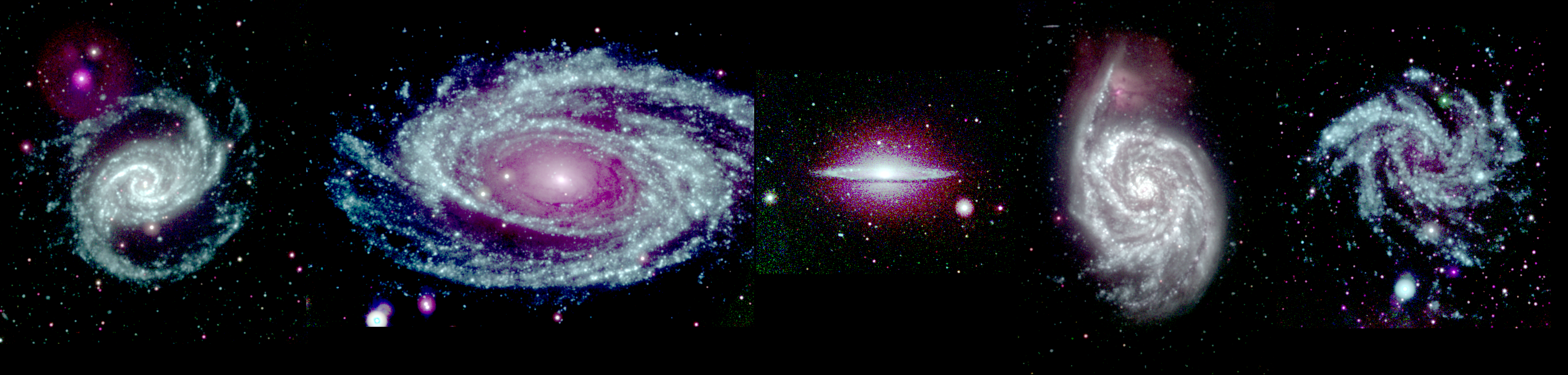}
    \caption{False color images of select SINGS/KINGFISH galaxies using Swift UVOT (UVW2, UVM2, and UVW1 filters). From left to right, NGC 1566, NGC 3031 (M81), NGC 4594 (M104), NGC 5194 (M51), and NGC 6946.}
    \label{fig:prettypics}
\end{figure*}

Multiwavelength integrated light photometry of the SINGS/KINGFISH galaxies was presented in \citet{Dale2017}, covering the far-UV to the far-IR. In this work, we add photometry from the three near-UV filters of the {\it Swift}/UVOT to the combined SINGS/KINGFISH sample. Here, we summarize the existing photometry and other measurements used in our SED fitting and discuss the raw {\it Swift}/UVOT data. We refer readers to \citet{Dale2017} for a more detailed discussion on the archival photometry.

\subsection{Archival Photometry}

We study the combined SINGS/KINGFISH sample of 79 nearby ($D_L < 30$ Mpc) galaxies, which have broadband coverage from the far-UV to the far-IR \citep{SINGS_2003PASP, KINGFISH_2011PASP..123.1347K}. The galaxies in the sample span a wide range of galaxy properties, which is representative of the local galaxy population based on morphological type, luminosity, and the ratio of FIR to optical luminosity. However, the extreme ends of the luminosity function are not well represented in the SINGS/KINGFISH sample. 

\citet{Dale2017} is the latest published version of the integrated light photometry for these galaxies, correcting flux calibrations that were reported in previous works \citep{2007ApJ...655..863D, 2012ApJ...745...95D}. Photometry includes data from {\it GALEX}, ground based optical data from SDSS, 2MASS, {\it Spitzer} IRAC and MIPS, {\it WISE},  {\it Herschel} PACS and SPIRE, JCMT Scuba, and {\it Planck} HFI. In total, thirty bands from 0.154 to 850 $\mu$m are used from \citet{Dale2017}. This dataset has been used in other studies for SED fitting \citep{Noll_09, Hunt_SINGS_sed, decleir_thesis, 2020ApJ...889..150A, 2022ApJ...931...53D}.

\subsection{{\it Swift}/UVOT data}

We present {\it Swift}/UVOT photometry for 74 of the 79 SINGS/KINGFISH galaxies; the observations are summarized in Table~\ref{tab:exptime}. The median exposure times for the UVW2, UVM2, and UVW1 filters are 6688, 7918, 5539 seconds, respectively. False-color images using {\it Swift}/UVOT data of select SINGS/KINGFISH galaxies are displayed in Figure~\ref{fig:prettypics}. The reduction and processing of the UVOT images is described in  Section~\ref{sec:methods}. 

Photometry is excluded for five galaxies from the SINGS/KINGFISH sample: IC 342, NGC 5195 (M51b), NGC 1482, NGC 3077, and NGC 5408. IC 342 is a face-on spiral galaxy but suffers from high foreground extinction ($E(B-V) \sim 0.4$). NGC 5195 is the interacting satellite of M51 which makes separating its' own emission from that of M51 difficult. The fields for NGC 1482 and NGC 3077 contain nearby bright foreground stars that affect data quality. Lastly, NGC 5408 is removed due to a nearby bright star making it observable in only the UVM2 filter. These galaxies are not considered during our SED fitting. We include M51 (NGC 5194) in our sample since it is 40 times brighter in {\it GALEX} NUV than the interacting NGC 5195 so contamination by its satellite is minimal.

We present the measured UVOT fluxes for all galaxies despite potential contamination due to non-thermal emission from Active Galactic Nuclei (AGN). Unobscured AGN could have significant UV flux which would impact our attenuation law measurement. Additionally, obscured AGN can have a significant effect in the mid-IR which could bias our estimate of stellar mass. We discuss identifying potential AGN in \S\ref{sec:SED_fitting}.

The addition of these near-UV bands to the archival data is paramount for constraining the shape of the dust attenuation law. Importantly, the {\it Swift}/UVOT filters straddle the UV bump making them uniquely sensitive to bump strength (see Figure~\ref{fig:dustlaws} and Table~\ref{tab:exptime} for filter transmission curve information); combined with the other UV filters which constrain the slope, the inclusion of UVOT photometry greatly improves the sampling of the UV SED where the effects of dust are strongest and change rapidly with wavelength. Specifically, the data used here involve 6 filters in the UV; three from \citet{Dale2017} ({\it GALEX} FUV at 0.154 $\mu$m, NUV at 0.234 $\mu$m, and SDSS $u$ at 0.35 $\mu$m) and three from {\it Swift}/UVOT (UVW2 at 0.214 $\mu$m, UVM2 at 0.227 $\mu$m, UVW1 at 0.269 $\mu$m). This dense UV sampling, combined with existing photometry that samples the optical stellar emission and the thermal dust emission in the far-IR, is imperative in order to accurately model the effects of dust.

\subsection{Emission Line Fluxes}

In addition to photometry, we use the observed H$\alpha$ and H$\beta$ fluxes from \citet{emission_line} to help constrain the recent star formation rate. The observed ratio of these emission line fluxes, when compared to the theoretical value of 2.86 in the absence of dust, yields an important estimate of the amount of reddening affecting the ionized gas around young stars \citep{2006agna.book.....O}. 

In addition to these emission lines providing information about extinction properties, Balmer emission lines trace recent star formation in the last 10 Myr as they directly trace the ionizing radiation from young stars. \citep{1983ApJ...272...54K, 1998ARA&A..36..189K}. These emission line fluxes are used as additional inputs to the SED fitting and help constrain important parameters such the dust law and recent star formation rate.

\subsection{Metallicity measurement for SED fitting}

We use metallicity information from the literature to guide our treatment of the stellar metallicity in the SED fits, specifically the nebular gas-phase oxygen abundance. The nebular oxygen abundances are measured for these galaxies using strong line methods in \citet{2010ApJS..190..233M}. Measured line fluxes were used to determine the characteristic nebular oxygen abundances using the theoretical relationships of \citet{KK04} \citep[see characteristic, globally averaged abundances given in Table 9 of][]{2010ApJS..190..233M}.

These oxygen abundances were converted to a metallicity measurement, assuming a solar abundance pattern, using 

\[    \log(Z/Z_\odot) = 12+\log(O/H) - 8.69, \]

where 8.69 is the solar photospheric oxygen abundance and $Z_\odot = 0.0134$ is the solar metallicity \citep{2009ARA&A..47..481A}. The stellar metallicity was estimated from the measured oxygen abundance for 52 of our 74 galaxies. Metallicity estimates for some of these galaxies have also been reported in \citet{2014A&A...563A..31R} and \citet{2019A&A...623A...5D}. However as neither contain metallicities for the entire SINGS/KINGFISH sample, we use the values quoted in \citet{2010ApJS..190..233M}. 

Since not all of the galaxies in our sample have measured oxygen abundances, we use the B-band luminosity-metallicity relationship values derived and quoted in  \citet{2010ApJS..190..233M}. They determined this empirical relationship using the measured spectroscopic abundances and B band luminosity. The metallicity of 20 galaxies was determined using the measured B band luminosity, which has been shown to hold over a wide range of luminosity stellar masses \citep[see][and references therein, about the L-Z relationship]{Lee_2006}. We note that the luminosity-metallicity relationship has a large scatter up to 0.2 dex and could potentially be biased due to the imbalance in the sample between early and late type galaxies. We use these values despite this to remain self-consistent. Lastly, we assumed a solar metallicity for the remaining 2 galaxies for which no other constraints were available. Note that this means we are assuming the gas-phase and stellar metallicities to be equal for the purposes of our SED fitting.

By fixing the metallicity, we aim to better estimate the age of the stellar populations and the dust effects. However, metallicities and chemical enrichment are sources of large uncertainty in galaxy studies. Specifically, there is uncertainty in converting from measured line fluxes to a metallicity measurement \citep{2019ARA&A..57..511K}. Additionally, one implicit assumption is that the solar abundance pattern is ubiquitous, which may lead to inaccurate metallicity estimates.

\begin{deluxetable}{c|ccc}
\tablecaption{ Summary of {\it Swift}/UVOT observations \label{tab:exptime}}
\tabletypesize{\scriptsize}
\tablehead{
\nocolhead{} & \colhead{UVW2} & \colhead{UVM2} & \colhead{UVW1} }
\startdata
Central wavelength & 2140.3 \AA & 2272.7 \AA & 2688.5 \AA \\
FWHM & 584.89 \AA & 527.13 \AA & 656.60 \AA \\
Median exposure time & 6689 s & 7918 s & 5539 s
\enddata
\tablecomments{Central wavelengths and FWHM values taken from SVO Filter Profile Service. The UVW1 and UVW2 filters suffer from a red leak which has been described in \citet{2010ApJ...721.1608B, 2011AIPC.1358..373B, siegel2014}.}

\end{deluxetable}

\section{Data Processing} \label{sec:methods}

In this section, we describe the process used to reduce the UVOT data and extract integrated flux measurements for the SINGS/KINGFISH galaxies. The raw data were downloaded from the High Energy Astrophysics Science Archive Research Center (HEASARC)\footnote{\url{https://heasarc.gsfc.nasa.gov/cgi-bin/W3Browse/swift.pl}}. All available UVOT data for the SINGS/KINGFISH galaxies were downloaded as of December 2021. Only raw images that were binned 2$\times$2 and properly aspect corrected were used. The code used to reduce the data and perform photometry can be found on GitHub\footnote{\url{https://github.com/UVOT-data-analysis}}. Description of a similar analysis of UVOT data for galaxies can be found in \citet{mallory} and \citet{Hagen_SMC}. The data reduction pipeline used here is similar to an independent code called DRESSCode\footnote{\url{https://spacetelescope.github.io/DRESSCode/}} (Decleir et al., in prep.), used in \citet{Decleir2019} \& \citet{decleir_thesis}.

\subsection{ {\it Swift}/UVOT data reduction}

Observations of these galaxies were taken as part of a fill-in program from 2005 to 2021. The targeted exposure time for each UVOT filter was 5 ks. Due to the orbit of {\it Swift}, the maximum, continuous observation length is 1800 seconds, so multiple exposures are required in order to achieve the desired depth. Therefore, we stack individual observations for both the sky count images and the exposure maps. Following the procedure laid out in \citet{siegel2014}, the images are corrected for detector issues such as large scale sensitivity variations and bad pixels. 

Then, we correct for time-dependent throughput loss affecting the sensitivity of UVOT. For this, we use the latest correction as of September 2020\footnote{\url{https://heasarc.gsfc.nasa.gov/docs/heasarc/caldb/swift/docs/uvot/uvotcaldb_throughput_06.pdf}}. For data taken 14 years into the mission (2019), the sensitivity loss is 34\%, 25\%, and 26\% in the UVW2, UVM2, UVW1 filters respectively. We correct each observation individually prior to stacking. 

Since multiple observations are being combined into one image, it is possible that differences in the background will impact the final science images. This is particularly important for the UVW1 and UVW2 filters which have significant red leaks leading to higher backgrounds \citep{2010ApJ...721.1608B, 2011AIPC.1358..373B}. The background can vary due to the relative position of the Moon or Sun when the observation is taken. In order to mitigate this risk, we combine individual observations in two different ways to test if a varying background will affect our photometry. Our first method is stacking the data without correcting for the individual background of each snapshot. The second method corrects the background level of each image before combining.

When correcting each individual snapshot, a sky region which was observed by each snapshot is determined. Within that sky region, a biweight of the sky pixel values is calculated for each observation. The minimum biweight value is used as a reference. In the rest of the observations, we subtract off the counts needed to have the sky area biweight match the reference. We refer to this as an offset correction.

Since UVOT is a photon counting instrument, coincidence loss can occur and the correction is given for point sources \citep{2008MNRAS.383..627P, 2010MNRAS.406.1687B}. However, there is not a standard correction for extended objects. \citet{hoversten} excluded regions above a certain count rate but found that it was only an issue for some individual star forming regions. As demonstrated in \citet{Decleir2019}, correcting for coincidence loss is important for studies on resolved scales, especially in star forming regions with high count rates. 

Other studies of nearby galaxies have noted that this correction is smaller than the deadtime correction so coincidence loss can be safely ignored, as it is not a significant source of error \citep{mallory}. We verified that our integrated photometry is not strongly affected by coincidence loss by comparing the UVM2 and {\it GALEX} NUV magnitudes, which agree when accounting for color differences due to the relative filter transmission curve shapes.

\subsection{UVOT photometry}

Once the data have been processed and stacked together, photometry is done to measure the integrated light from these galaxies. The apertures from \citet{Dale2017} are used and prior to photometry, foreground stars are masked using the {\it GALEX} NUV resolution masks from \citet{z0mgs}. 

When measuring the photometry, we subtract the mean background determined using an annulus around the galaxy where the outer edge of the annulus was $\sim3$ arcminutes larger than the \citet{Dale2017} apertures. As mentioned above, we measure photometry using two stacking methods.

\begin{figure*}
    \centering
    \includegraphics[width=\textwidth]{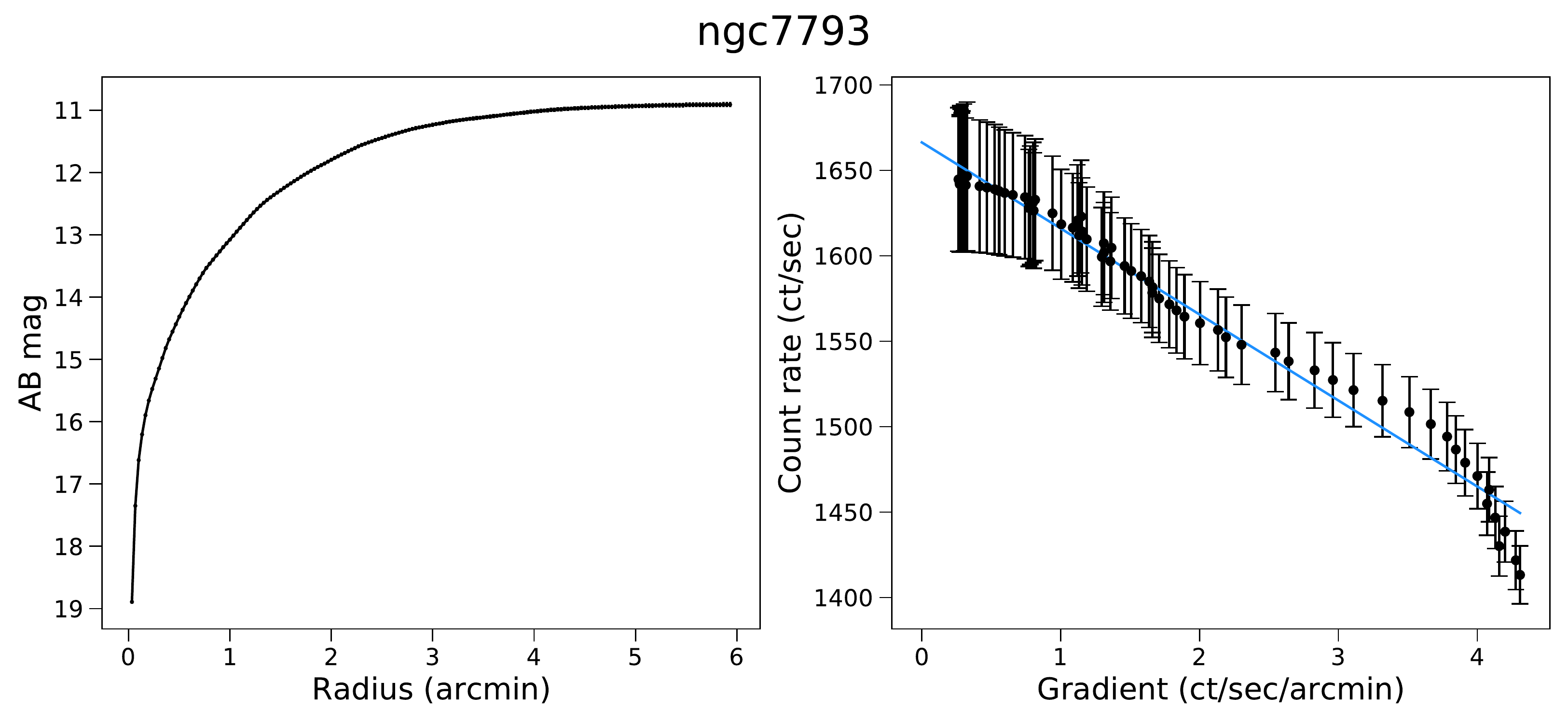}
    \caption{ {\it Swift}/UVOT UVW1 photometry measurement for NGC 7793.  {\it Left:} Source brightness as a function of aperture size, which clearly asymptotes at large radii. {\it Right:} Count rate as a function of rate of change of flux accumulation. Following \citet{GildePaz}, we calculate the asymptotic flux by determining the count rate where the rate of change of the count rate with respect to radius equals zero. This calculation is done only at large radii where very little flux is still being accumulated as the aperture increases. The blue line is the best fit line, where the intercept is used as the asymptotic flux measurement. }
    \label{fig:asymptotic_mag}
\end{figure*}

The photometry is measured in a similar fashion as was done with {\it GALEX} data in \citet{GildePaz}, \citet{2009ApJ...703.1569M} \& \citet{GALEX_SB}. Elliptical apertures of increasing size, up to the aperture used in \citet{Dale2017}, are used to determine the count rate accumulated as a function of radius. This brightness profile is extrapolated to the asymptotic count rate. An example of a brightness profile is shown in Figure~\ref{fig:asymptotic_mag}. We convert the count rate to AB magnitudes using the zero-point offsets from \citet{2011AIPC.1358..373B} and convert to $f_\nu$ according to \citet{1983ApJ...266..713O}. Since the zero-points were calibrated for 5 arcsecond apertures, we must utilize the curve of growth from \citet{2010MNRAS.406.1687B} to account for our much larger apertures. This correction means the measured flux densities using the zero-point offsets from \citet{2011AIPC.1358..373B} must be divided by 1.1279, 1.1777, and 1.1567 for the UVW2, UVM2, and UVW1 filters respectively \citep[see][]{Decleir2019, decleir_thesis}.

The angular extent of M101 is larger than the UVOT field of view, and is still not fully covered by our final mosaic. Using the aperture from \citet{Dale2017} would require a significant aperture correction. Instead, we perform the UVOT photometry for M101 using a custom aperture (center at (RA, Dec) of (210.8417, 54.3556), major axis: 1400" , minor axis: 1080", PA: 50\degree). The {\it GALEX} NUV and UVOT UVM2 filters are centered at approximately the same wavelength so we compare the UVM2-NUV color for M101 to the rest of the sample. M101's UVM2-NUV color is 0.153 which is in line with the rest of the sample (mean UVM2-NUV color is 0.1). Therefore, we keep M101 in our sample and use this custom aperture.

We find that the photometry measured for both stacking methods described above agree with each other. However, the non-offset corrected images occasionally produce photometry that have unrealistically large error bars for a handful of galaxies (i.e. photometric errors are consistent with no flux/are only upper limits). As a result, the offset corrected images are used in our SED fitting analysis and reported here. The correlation coefficients for the offset and nonoffset images are greater than 0.998 for each of the three NUV filters. The errors on the offset corrected images are smaller compared to the non-offset corrected images, likely due to a better background subtraction. 

The flux densities for the SINGS/KINGFISH galaxies in the {\it Swift}/UVOT filters are provided in Table~\ref{tab:fluxes}. The presented fluxes are not corrected for foreground MW dust extinction or internal attenuation and only represent the observed values. The errors reported are the quadrature sum of the zero-point errors and the statistical error on the intercept, representing the asymptotic value as shown in Figure~\ref{fig:asymptotic_mag}.

\startlongtable
\begin{deluxetable*}{ccccccc}

\tablecaption{{\it Swift} UVOT uncorrected fluxes of SINGS/KINGFISH galaxies studied here \label{tab:fluxes}}

\tabletypesize{\scriptsize}
\tablehead{\\
Galaxy Name & UVW2 & UVW2 error & UVM2 & UVM2 error & UVW1 & UVW1 error \\
\nocolhead{} & 2140.3 \AA & \nocolhead{} & 2272.7 \AA & \nocolhead{} & 2688.5 \AA & \nocolhead{} 
}
\startdata 
NGC 24 &  10.376 &     0.320 &  10.223 &     0.321 &  13.857 &     0.686 \\
NGC 337 &   8.384 &     0.232 &   8.456 &     0.234 &  12.217 &     0.339 \\
NGC 584 &   1.425 &     0.387 &   1.063 &     0.083 &   6.194 &     0.676 \\
NGC 628 &  61.517 &     1.715 &  54.594 &     1.574 & 105.283 &     4.944 \\
NGC 855 &   1.713 &     0.052 &   1.667 &     0.047 &   2.842 &     0.276 \\
NGC 925 &  38.628 &     1.074 &  35.093 &     0.992 &  52.320 &     1.473 \\
NGC 1097 &  36.849 &     1.033 &  34.097 &     0.949 &  56.676 &     1.645 \\
NGC 1266 &   0.204 &     0.189 &   0.162 &     0.030 &   0.682 &     0.396 \\
NGC 1291 &  17.406 &     0.768 &  14.310 &     0.817 &  46.859 &     1.651 \\
NGC 1316\dag &  10.558 &     0.461 &   8.531 &     0.557 &  31.115 &     1.034 \\
NGC 1377\dag &   0.357 &     0.051 &   0.357 &     0.022 &   1.463 &     0.639 \\
NGC 1404\dag &   4.465 &     0.651 &   2.641 &     0.102 &  13.662 &     2.141 \\
NGC 1512 &  21.357 &     2.157 &  19.397 &     0.838 &  33.049 &     4.759 \\
NGC 1566 &  57.129 &     1.740 &  54.886 &     1.542 &  72.494 &     2.715 \\
NGC 1705 &  11.353 &     0.321 &  12.536 &     0.357 &  12.373 &     0.409 \\
NGC 2146 &   4.865 &     0.292 &   4.260 &     0.131 &   9.961 &     0.909 \\
NGC 2403 & 219.067 &     6.257 & 201.538 &     5.857 & 288.300 &     9.557 \\
UGC 4305 &  35.225 &     1.015 &  33.089 &     0.928 &  39.583 &     1.265 \\
 M81 dwA &   0.362 &     0.071 &   0.352 &     0.029 &   0.350 &     0.164 \\
 DDO 53 &   2.222 &     0.097 &   1.926 &     0.060 &   2.935 &     1.599 \\
NGC 2798\dag &   2.988 &     0.131 &   3.005 &     0.153 &   4.837 &     0.347 \\
NGC 2841 &  17.421 &     0.494 &  14.614 &     0.405 &  32.615 &     1.068 \\
NGC 2915 &   2.317 &     0.068 &   1.866 &     0.052 &   3.363 &     0.130 \\
UGC 5139 &   3.782 &     0.125 &   3.411 &     0.111 &   4.473 &     0.233 \\
NGC 2976 &  19.285 &     0.558 &  16.907 &     0.470 &  30.699 &     0.920 \\
NGC 3049 &   3.640 &     0.115 &   3.425 &     0.096 &   4.492 &     0.567 \\
NGC 3031 & 136.928 &     4.318 & 104.358 &     9.895 & 235.394 &     7.452 \\
NGC 3034 &  29.820 &     0.874 &  21.663 &     0.610 &  66.849 &     1.880 \\
UGC 5336 &   2.319 &     0.236 &   2.218 &     0.142 &   2.912 &     0.365 \\
UGC 5423 &   0.601 &     0.039 &   0.525 &     0.021 &   0.688 &     0.155 \\
NGC 3190 &   1.485 &     0.462 &   0.887 &     0.115 &   5.141 &     1.656 \\
NGC 3184 &  42.285 &     1.200 &  40.768 &     1.128 &  62.845 &     2.617 \\
NGC 3198 &  24.807 &     0.759 &  23.511 &     0.740 &  35.021 &     1.209 \\
 IC 2574 &  42.366 &     1.341 &  39.483 &     1.150 &  52.834 &     2.034 \\
NGC 3265 &   0.857 &     0.043 &   0.834 &     0.024 &   1.857 &     0.252 \\
  Mrk 33\dag &   4.207 &     0.117 &   4.482 &     0.124 &   4.834 &     0.146 \\
NGC 3351 &  23.403 &     0.660 &  23.834 &     0.662 &  39.509 &     1.164 \\
NGC 3521 &  31.315 &     1.124 &  26.776 &     1.061 &  60.890 &     2.472 \\
NGC 3621\dag &  57.720 &     1.912 &  54.442 &     1.804 &  85.384 &     2.757 \\
NGC 3627 &  48.288 &     1.365 &  44.922 &     1.242 &  84.486 &     2.424 \\
NGC 3773 &   3.607 &     0.100 &   3.786 &     0.105 &   4.175 &     0.118 \\
NGC 3938 &  33.375 &     0.925 &  32.346 &     0.894 &  46.132 &     1.306 \\
NGC 4125 &   3.048 &     0.433 &   1.534 &     0.073 &   9.392 &     0.858 \\
NGC 4236 &  77.504 &     3.278 &  78.517 &     3.140 &  94.649 &     3.396 \\
NGC 4254\dag &  42.420 &     1.173 &  41.748 &     1.154 &  62.344 &     1.727 \\
NGC 4321 &  44.554 &     1.233 &  43.781 &     1.211 &  65.607 &     1.833 \\
NGC 4450\dag &   5.096 &     0.247 &   3.767 &     0.141 &  13.467 &     1.382 \\
NGC 4536 &  19.479 &     0.565 &  18.310 &     0.533 &  29.962 &     0.900 \\
NGC 4552 &   4.023 &     0.234 &   2.095 &     0.080 &  11.772 &     0.674 \\
NGC 4559 &  54.184 &     1.566 &  52.390 &     1.514 &  67.073 &     1.943 \\
NGC 4569\dag &  14.278 &     1.042 &  12.579 &     0.468 &  28.354 &     1.713 \\
NGC 4579\dag &   8.549 &     0.270 &   7.922 &     0.220 &  20.476 &     1.183 \\
NGC 4594 &  19.321 &     1.231 &   9.835 &     2.415 &  52.179 &     2.044 \\
NGC 4625 &   5.858 &     0.490 &   5.774 &     0.222 &   7.107 &     1.009 \\
NGC 4631 & 111.045 &     3.096 & 110.540 &     3.075 & 137.233 &     3.849 \\
NGC 4725 &  28.599 &     0.907 &  25.215 &     0.739 &  46.464 &     1.779 \\
NGC 4736 &  84.098 &     2.926 &  77.745 &     2.302 & 136.350 &     4.250 \\
 DDO 154 &   3.833 &     0.139 &   3.572 &     0.102 &   4.351 &     0.268 \\
NGC 4826 &  25.358 &     0.724 &  21.977 &     0.608 &  55.537 &     1.595 \\
 DDO 165 &   6.384 &     0.182 &   6.200 &     0.172 &   7.326 &     0.252 \\
NGC 5033\dag &  23.409 &     0.728 &  21.825 &     0.659 &  35.929 &     1.248 \\
NGC 5055 &  63.282 &     1.839 &  58.360 &     1.776 & 112.493 &     3.317 \\
NGC 5194 & 161.221 &     4.489 & 157.385 &     4.353 & 233.038 &     6.459 \\
NGC 5398 &   5.878 &     0.177 &   6.311 &     0.203 &   7.675 &     0.322 \\
   M101 & 337.615 &    12.774 & 336.040 &    19.015 & 414.825 &    13.823 \\
NGC 5474 &  25.679 &     0.752 &  24.845 &     0.694 &  30.176 &     1.030 \\
NGC 5713 &   7.414 &     0.299 &   6.897 &     0.280 &  12.063 &     0.705 \\
NGC 5866 &   4.120 &     0.590 &   3.068 &     0.089 &  14.319 &     1.299 \\
 IC 4710 &   8.116 &     0.299 &   7.657 &     0.260 &  10.269 &     0.447 \\
NGC 6822 &  79.593 &     6.589 &  66.859 &    11.812 & 144.101 &    12.779 \\
NGC 6946 &  40.718 &     1.386 &  28.905 &     0.872 &  80.830 &     2.579 \\
NGC 7331 &  15.145 &     0.433 &  13.009 &     0.370 &  31.514 &     0.938 \\
NGC 7552 &   9.787 &     0.275 &   9.390 &     0.260 &  15.143 &     0.474 \\
NGC 7793 & 132.918 &     3.674 & 130.131 &     3.597 & 160.667 &     4.454
\enddata
\tablecomments{Uncorrected fluxes are reported in mJy. The flux reported here for M101 uses a different aperture than \citet{Dale2017}. Galaxies marked with \dag\ are flagged due to probable AGN contamination. }
\end{deluxetable*}

\subsection{Correcting for MW foreground extinction}

Once the observed fluxes of the galaxies had been measured, the {\it Swift}/UVOT fluxes were combined with the \citet{Dale2017} photometry. Prior to SED fitting, we must correct our photometry for foreground MW extinction. Since this work focuses on the effects of internal dust attenuation, care must be taken in correcting for MW foreground extinction. 

To correct for foreground extinction, the \citet{F99} average MW extinction law with $R_V=3.1$ is assumed. The color excesses along the line of sight to these galaxies are taken from \citet{2011ApJ...737..103S} and compiled in \citet{Dale2017}. To calculate the true correction, the true spectral slope over the wavelength range covered by the filter transmission curves is needed; the spectrum and filter transmission weighted central wavelength is then used to get the absolute extinction from a chosen extinction law. In practice the spectrum is not known a priori. An additional wrinkle for large extended sources is that MW foreground dust could be nonuniform across the source leading to differential reddening which is not taken into account here.

In our correction for MW foreground extinction, we assume a flat spectrum across the filter of interest to calculate the central wavelength, meaning we calculate the transmission weighted central wavelength. \citet{mallory} test the effects of extinction law choice on {\it Swift} UVOT colors using different spectra from Starburst99 and find that the difference is on the order of a few percent for the highest $E(B-V)\sim0.1$ object in their sample \citep{1999ApJS..123....3L}. Here, the median reddening for our sample is $E(B-V)=0.027$. Thus, assuming a flat spectrum and the \citet{F99} law for MW foreground extinction introduces a negligible error on the flux measurements. Similar tests were done in \citet{Decleir2019} and found that assuming a flat spectrum is negligible compared to other sources of error.

\section{SED fitting} \label{sec:SED_fitting}

\begin{figure}
    \centering
    \includegraphics[width=0.48\textwidth]{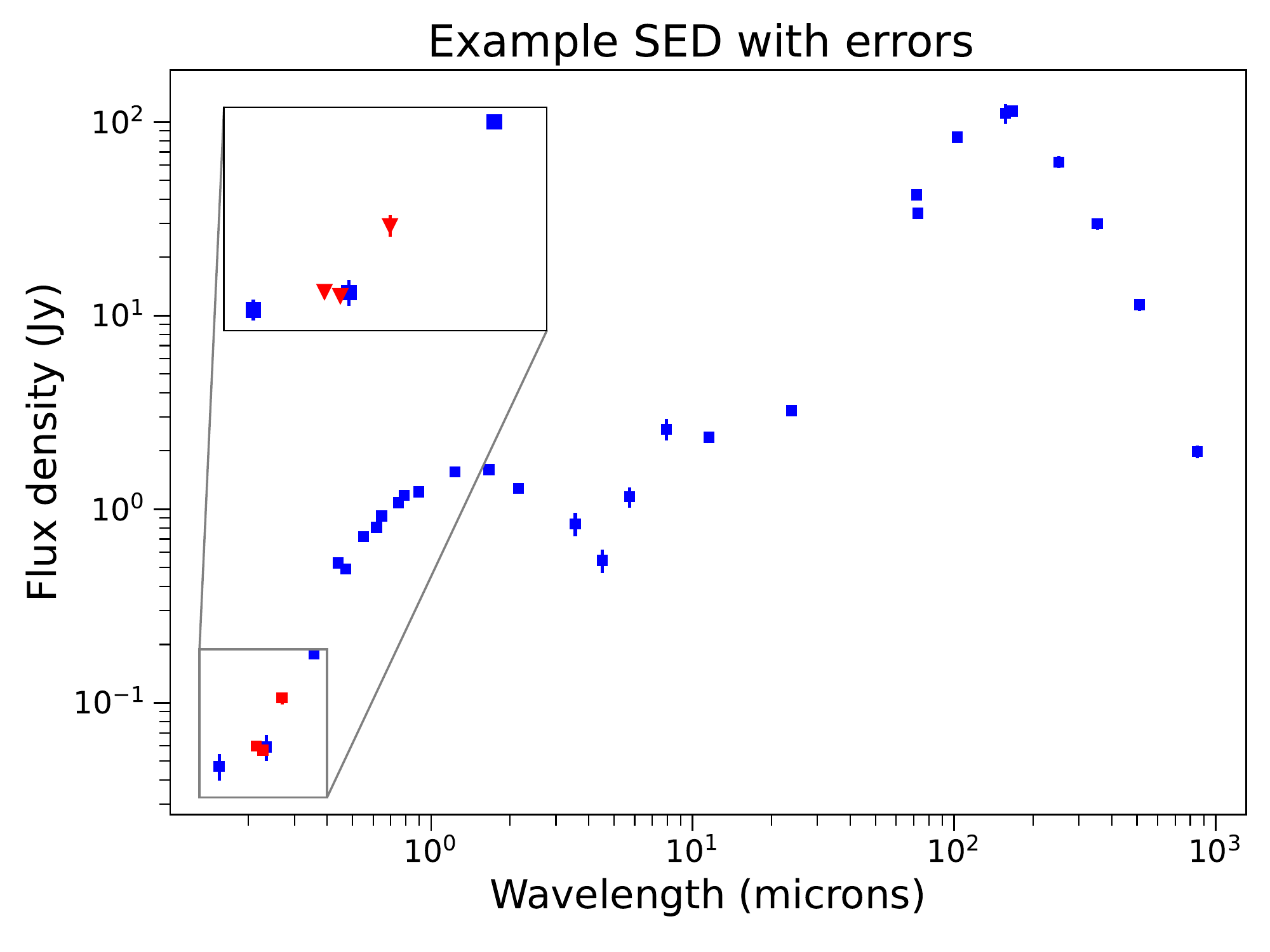}
    \caption{An example SED of NGC 628 (M74) showing the flux observed in each bandpass. The inset in the upper left shows the additional UV coverage provided by {\it Swift}/UVOT in red. The error bars on the fluxes are shown but are negligible in most cases. These data are then used as the inputs for our SED fitting.}
    \label{fig:example_sed}
\end{figure}

After combining the photometry and correcting for the effects of MW dust extinction, the full panchromatic SEDs were constructed for each galaxy; an example is shown in Figure~\ref{fig:example_sed}. In Table~\ref{tab:filtcount}, we report the number of galaxies observed in each filter to show our wavelength coverage. In total, our data includes 30 bands of archival photometry, three bands of new UVOT photometry, H$\alpha$ \& H$\beta$ fluxes, and metallicity measurements via oxygen lines \citep{emission_line, 2010ApJS..190..233M, Dale2017}. These data are the inputs for our SED fitting using \verb|MCSED| \citep{2020ApJ...899....7B}. While many different SED fitting codes exist including CIGALE, GRASIL, MAGPHYS, BAGPIPES, and Prospector \citep{Noll_09, 2019A&A...622A.103B, GRASIL, MAGPHYS, 2018MNRAS.480.4379C, 2017ApJ...837..170L, 2021ApJS..254...22J}, we use \verb|MCSED| due to its flexibility, allowing us to tailor it to our specific use case. It is important to be explicit with the assumptions made during SED fitting as different codes handle components like dust emission and metallicity effects differently as well as assume different parameterizations of the star formation history and the initial mass function. To illustrate the difficulty of SED fitting, \citet{2022MNRAS.514.5706J} examined the effects of using various stellar population and dust emission models and found considerable scatter. In this Section, we present and justify the choices made here as well as examine what can be tested by varying our assumptions.

\begin{deluxetable}{cc}
\tablecaption{Number of galaxies\\ observed in each filter \label{tab:filtcount}}
\tablehead{\\Filter & Count}
\startdata
{\it GALEX} FUV &    70 \\
{\it Swift} UVW2 &    74 \\ 
{\it GALEX} NUV &    73 \\
{\it Swift} UVM2 &    74 \\      
{\it Swift} UVW1 &    74 \\     
SDSS $u$ &    49 \\ 
B &    74 \\
SDSS $g$ &    49 \\ 
V &    74 \\          
SDSS $r$ &    49 \\  
$R_c$ &    69 \\        
SDSS $i$ &    49 \\  
$I_c$ &    71 \\         
SDSS $z$ &    49 \\    
2MASS $J$ &    74 \\
2MASS $H$ &    74 \\
2MASS $K_s$ &    74 \\
IRAC 3.6 $\mu$m &    74 \\ 
IRAC 4.5 $\mu$m &    74 \\ 
IRAC 5.7 $\mu$m &    74 \\  
IRAC 8.0 $\mu$m &    74 \\  
{\it WISE} W3  &    74 \\     
MIPS 24 $\mu$m &    74 \\    
MIPS 72 $\mu$m &    74 \\   
PACS 72 $\mu$m &    62 \\
PACS 103 $\mu$m &    57 \\ 
MIPS 157 $\mu$m &    74 \\    
PACS 167 $\mu$m &    62 \\ 
SPIRE 252 $\mu$m &    62 \\
SPIRE 353 $\mu$m &    62 \\
SPIRE 511 $\mu$m &    62 \\   
{\it Planck} 850 $\mu$m &    36 \\     
Scuba 850 $\mu$m &    25 \\    
H$\alpha$ emission line &    54 \\  
H$\beta$ emission line &    54 \\
\enddata
\end{deluxetable}

\subsection{Assumptions and parameterizations}

The most pertinent assumptions made here involve the dust attenuation law and the star formation history, as we are interested in characterizing the variety of dust attenuation curves and the SFH is relatively unconstrained by the data. Other assumptions like the method for stellar population synthesis and the dust emission model are also described.

\subsubsection{Attenuation law} \label{atten}

Based on the variety of dust laws seen between the Magellanic Clouds, MW, and starburst galaxies, a flexible parameterization is needed to describe the attenuation law. Using the \citet{C00} starburst law as a base, \citet{Noll_09} introduced the following parameterization which was adapted in \citet{Kriek+Conroy}:
\begin{equation}
    \frac{A(\lambda)}{A(V)} = \frac{1}{R_{V, Cal}} \left(\kappa'(\lambda)  + D(\lambda, E_b)\right) \left(\frac{\lambda}{\lambda_V}\right)^\delta  
    \label{eq:noll_law}
\end{equation}

where $A(\lambda)$ is the extinction in magnitudes at wavelength $\lambda$, $A(V)$ is the visual extinction (at 5500 \AA), $\kappa'(\lambda)$ is the starburst attenuation law from \citet{C00}, $R_{V, Cal}=4.05$ is the ratio of visual extinction to color excess for the starburst law, $\delta$ represents a power law deviation relative to the starburst law, and $D(\lambda, E_b)$ is a Lorentzian-like Drude profile where $E_b$ sets the bump strength \citep{1986ApJ...307..286F}. Note that for this formulation, $\delta$ and $E_b$ are not independent and the power law deviation from the starburst law will affect the bump strength.

Other works keep the bump independent from the slope by applying the power law only to the \citet{C00} law and then adding in the Drude profile. For example, this is the parameterization used in the current version of CIGALE \citep{2019A&A...622A.103B}. The lack of a consistent usage in the literature makes comparisons problematic. In the form assumed here, since $\delta$ and $E_b$ are not independent, a relationship between the two might arise naturally without a physical meaning. However, there is theoretical backing \citep{2018ApJ...869...70N}  and observational evidence \citep[e.g.][]{Kriek+Conroy, Salim_2018, Decleir2019} for a physical relationship between the two.

The parameters $\delta$ and $E_b$, which describe the UV slope and the strength of the UV bump respectively, describe the departures from the well known starburst law and yield a rather simple and flexible formalism for describing the UV attenuation law. We assume the bump profile $D(\lambda, E_b)$ has a fixed central wavelength and width. \citet{1986ApJ...307..286F} found that the location of the bump was relatively constant but that the bump width was more variable. Here we adopt a central wavelength of 2175 \AA\ and a width of 350 \AA\, consistent with \citet{Kriek+Conroy}. A negative value of $\delta$ corresponds to a steeper attenuation law compared to the \citet{C00} starburst law, corresponding to $\delta=0$ (where a positive value yields a curve that is shallower than \citet{C00}). In this parameterization, the average MW extinction law is best described by $\delta\sim -0.1$ and $E_b\sim3$.

Measured dust curves like the SMC/LMC/SB curves can be derived using spectra. For most galaxies, only broadband photometry is available. As a result, the spectrally derived laws are able to capture more structure than those inferred using a handful of UV and optical photometric data points. Due to the complexity of spectrally derived dust laws, the two parameter \citet{Noll_09} law can struggle to fully describe them. For example, in order to describe the SMC law using the \citet{Noll_09} law, one must decide which features to accurately capture. The SMC law does not contain a dust bump and its slope is more complicated than a power law deviation from the \citet{C00} law. When trying to fit the \citet{Noll_09} by minimizing the residuals over UV and optical wavelengths, one finds a slope of $\delta=-0.478$, but with a negative bump strength, which is obviously not physically motivated.

Alternatively, \citet{2020ARA&A..58..529S} choose to describe the attenuation laws by focusing on the slope, $S$, defined as the ratio of attenuation at 1500 and 5500 \AA\, and the bump strength, $B$, defined as the ratio of the measured attenuation at 2175 \AA\ to the underlying baseline. They connect the \citet{Noll_09} parameter $\delta$ with $S$ but do not explicitly connect $E_b$ and $B$.  Using the \citet{2020ARA&A..58..529S} approach for the SMC law yields a slope $S=4.767$, which corresponds to $\delta=-0.5059$ and overestimates the amount of extinction at UV wavelengths. One could also try to best match the optical slope, which would underestimate extinction at UV wavelengths. 

None of the different approaches accurately reproduces the \citet{G03} SMC dust law over the entire wavelength range. The impreciseness in which we can define an accurate parametric dust attenuation law is an inherent weakness in describing the complexities seen in the closest galaxies. In summary, the two parameter attenuation law from \citet{Noll_09}, while allowing sufficient flexibility to approximate many different attenuation laws, can still struggle to reproduce the well-studied dust curves of the Magellanic clouds. It is unclear if an adequate alternative exists. 

A recent, promising alternative was suggested in \citet{2022ApJ...931...14L}. \citet{Kriek+Conroy} found a relationship between the slope $\delta$ and the bump strength $E_b$ for their sample of high redshift galaxies, making the \citet{Noll_09} law a single parameter model. \citet{2022ApJ...931...14L} assumes the $\delta - E_b$ relationship from \citet{Kriek+Conroy} and introduces a new parameter $f_{obscured}$ where only a fraction of the stellar light is obscured by dust, mimicking a more complex star-dust geometry than the typical screen model. This formulation of the attenuation law is not explored here but presents an interesting way to capture variations in the dust law, while remaining a simple two parameter model.

\subsubsection{Star formation history}\label{sec:sfh}

While we are primarily interested in the dust attenuation law, we want to test how different star formation history parameterizations affect the derived attenuation law. \citet{2019ApJ...873...44C} and \citet{2019ApJ...876....3L} studied the effects of the chosen SFH parameterization and found that a more flexible SFH with carefully chosen priors was best able to reproduce galaxy properties. Other studies like \citet{lower20} and  \citet{2022ApJ...935..146S} have further investigated the impact that the chosen SFH has recovering stellar mass and how best to fit quenched galaxies. This demonstrates that while the SFH is relatively unconstrained by the data, it can have a significant impact on the derived galaxy properties. 

\begin{figure}
    \centering
    \includegraphics[width=0.48\textwidth]{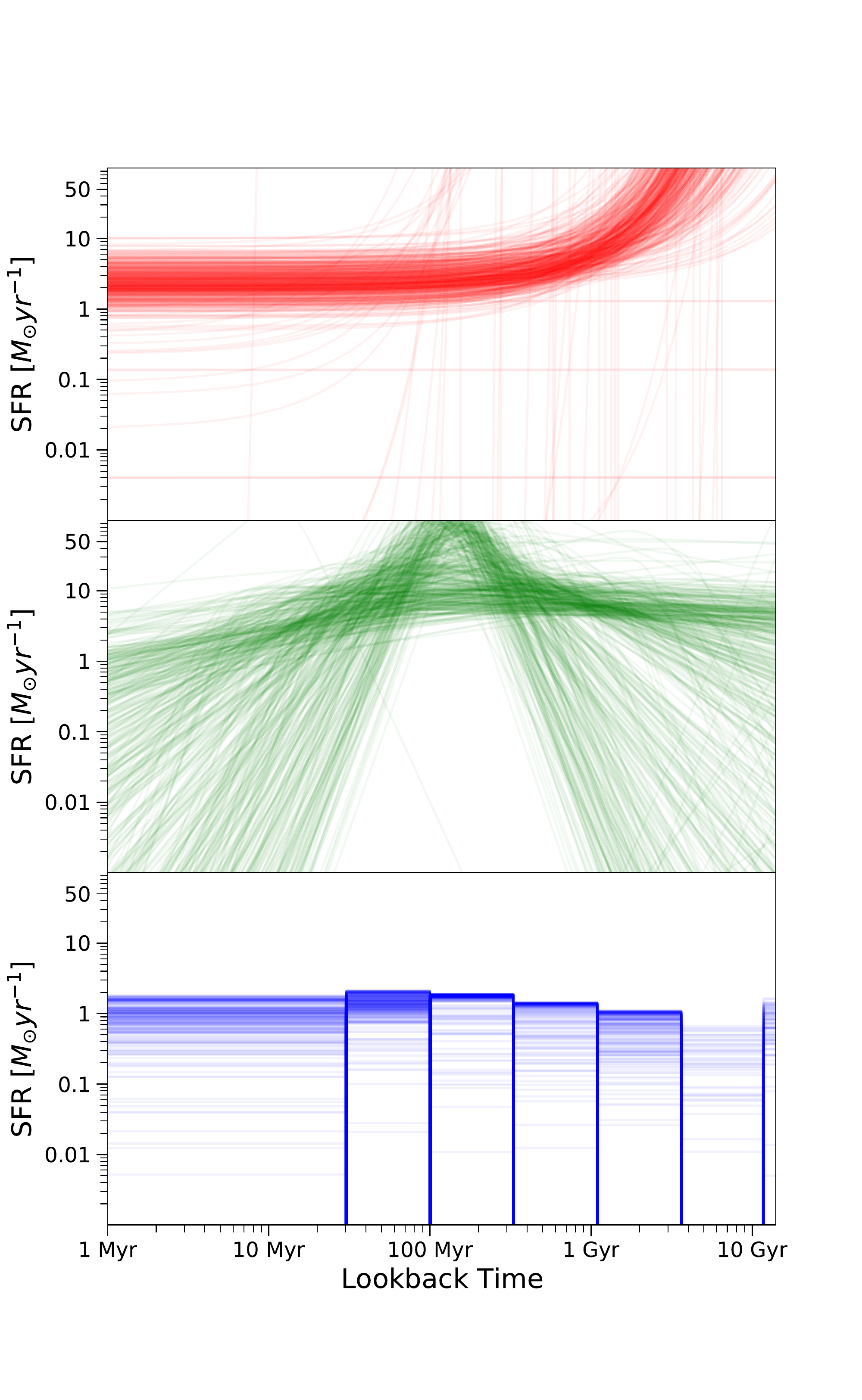}
    \caption{The three SFH parameterizations used during our fitting. From top to bottom, exponential (red), double power law (green), and piecewise (blue). This plot shows 500 realizations of the SFH for M101 from our SED fitting. The exponential and double power law forms are the most rigid in their assumed forms but are used due to the ease of computation. The piecewise SFH is ``non-parametric" in that no rigid shape is imposed on the SFH and is thought to be better at producing a more physically realistic SFH. However, it has the most parameters and therefore is more computationally demanding. The photometric data do not constrain the earliest epochs of star formation and these are more dependent on the priors and shape of the chosen SFH.} 
    \label{fig:sfh}
\end{figure}

Following \citet{2019ApJ...873...44C} and \citet{2019ApJ...876....3L}, we choose two parametric and a single more flexible ``non-parametric" SFH. Our two parametric formulations are a decaying exponential and a double power law. \citet{dpl_sfh} claim that the double power law SFH is more accurate than an exponential model and a good fit for individual galaxies. Similar to a double power law, \citet{2017A&A...608A..41C} argues for a right skewed Gaussian SFH. Our ``non-parametric" SFH is a piecewise function where a constant SFR is assumed in different age bins with discontinuities between age bins. Non-parametric is a misnomer since an explicit functional form is assumed but this parameterization allows the greatest flexibility. The location of the breaks are chosen to be physically motivated and occur at 30 Myr, 100 Myr, 330 Myr, 1.1 Gyr, 3.6 Gyr, and 11.7 Gyr to yield seven bins \citep{2006MNRAS.365...46O, 2019ApJ...876....3L}. The oldest five bins are logarithmically spaced while the two most recent bins are chosen to better model the youngest and bluest stellar populations. Figure~\ref{fig:sfh} shows the different SFH parameterizations used here. 

\subsubsection{Other assumptions}
 
To model our stellar populations, we use Flexible Stellar Population Synthesis (FSPS) and the Padova isochrones \citep{fsps_1, fsps_2, python_fsps, padova, padova2}. We use the dust emission models from \citet{DL07} and include excess attenuation for stars younger than 10 Myr following \citet{CF00} ($E(B-V)_{old} = 0.44 E(B-V)_{young}$). We do not assume energy balance and let the normalization of the FIR emission be a free parameter. Nebular emission is modelled using Cloudy with the underlying stellar populations from FSPS as an input and we assume a fixed ionization parameter $\log U = -3$ \citep{1998PASP..110..761F, 2017ApJ...840...44B, 2010ApJS..190..233M}. We assume a minimum flux uncertainty of 5\%. For the initial mass function, we assume a \citet{kroupa} IMF. 

As mentioned previously, we fix the metallicity based on gas phase oxygen abundances. Since the stellar libraries used by \verb|MCSED| only take certain discrete values for metallicity, the sample was binned to match the closest allowed values. The final choice of binning was $Z =$ 0.0016, 0.0025, 0.0039, 0.0061, 0.015, 0.019, 0.024, and 0.03. These bins had 1, 4, 3, 2, 8, 12, 13, 31 galaxies respectively. The construction of this sample leads to a paucity of low-mass, quiescent galaxies which could skew sample level trends. A larger statistical sample is needed to generalize about the galaxy population as a whole.

These assumptions are made about the physics of galaxies but we also must understand the statistical choices made for our SED fitting.

\begin{figure*}
    \centering
    \includegraphics[width=\textwidth]{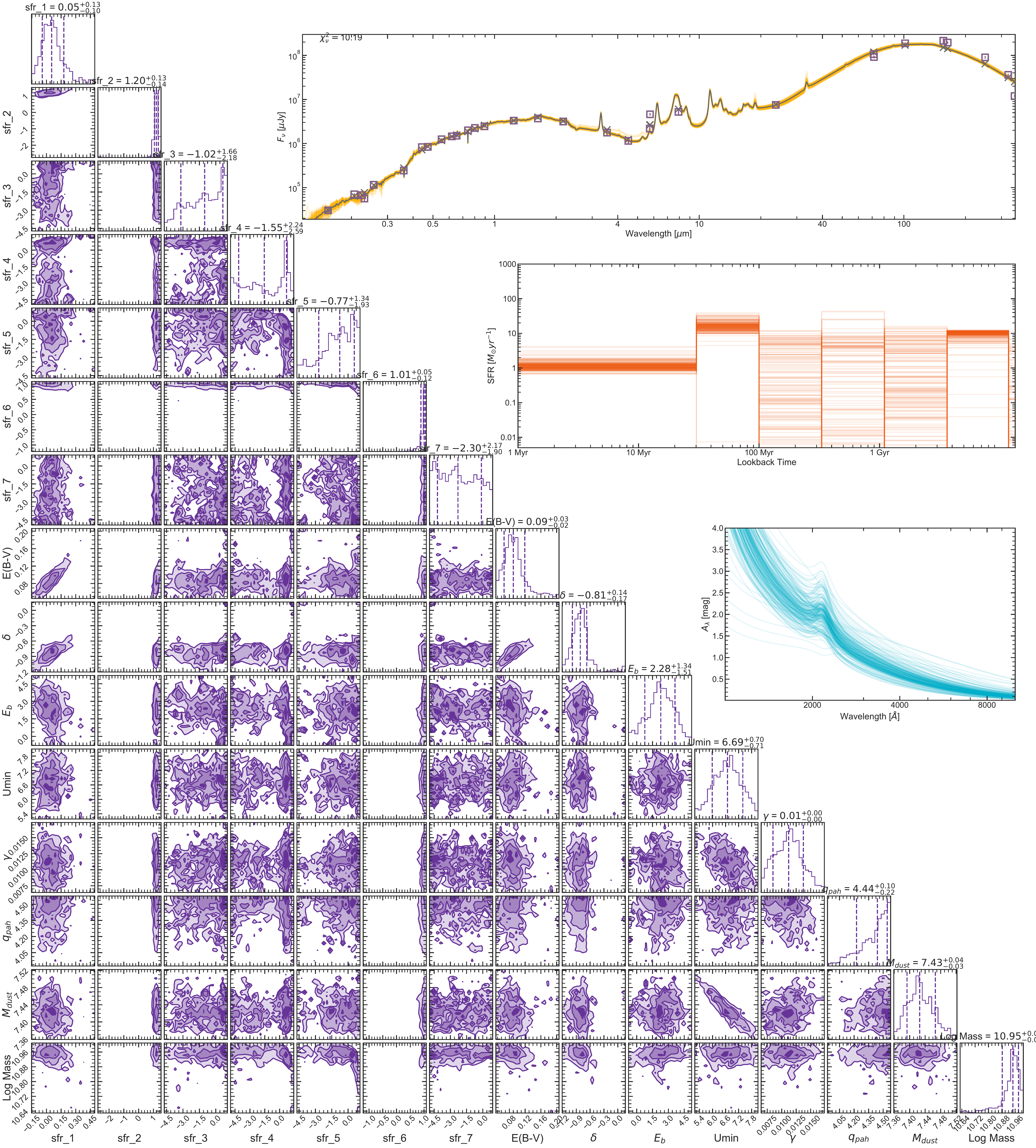}
    \caption{Example output corner plot for MCSED for NGC 3267 and the piecewise SFH \citep{corner}. This figure summarizes the parameter distributions. The posterior probability distributions of the different parameters can be seen in the lower left. The best fit spectrum, SFH, and attenuation law are shown in the upper right. This plot allows a quick check to see if the fit was successful in modelling the emission from the given galaxy and if there are any degeneracies between parameters. }
    \label{fig:triangle}
\end{figure*}

\subsection{Using MCSED}

The code for \verb|MCSED| can be accessed via GitHub\footnote{\url{https://github.com/wpb-astro/MCSED}} \citep{2020ApJ...899....7B}. Due to the flexibility allowed by this SED fitting code, we choose our set of assumptions and make small changes to best fit our science problem. 

MCSED uses an affine invariant ensemble sampler to explore parameter space and compare the measured photometry to the models described above \citep{2010CAMCS...5...65G}. This is done using the \verb|emcee| python package \citep{emcee}.  A total of 500 walkers are initialized and allowed to walk for 20,000 steps.

The priors used in \verb|MCSED| are uniform over a predefined region of parameter space. Table~\ref{tab:priors} contains the minimum and maximum values for priors of the different parameters in \verb|MCSED|. These values were chosen based on the defaults in \verb|MCSED| and what is physical. For example, the upper limit on $\delta$ is set to 0.5 since it is consistent with a flat attenuation law and allowing more positive slopes is nonphysical. 

\begin{table}[]
    \centering
    \begin{tabular}{|c|c|c|}
        \hline
        Parameter & Min & Max  \\\hline
        Dust attenuation & & \\\hline
        $E(B-V)$ & -0.05 & 0.75 \\
        $\delta$ & -1.6 & 0.5  \\
        $E_b$ & -1 & 6 \\\hline
        Dust emission & & \\\hline
        $U_{min}$ & 0.1 & 25 \\
        $\gamma$ & 0 & 1 \\
        $q_{PAH}$ & 0.47 & 4.58 \\
        $\log(M_{\rm dust})\ (M_\odot)$ & 4.5 & 10 \\\hline
        Exponential SFH & & \\\hline
        $\log$(SFR) (log($M_\odot$/yr)) & -7 & 5 \\
        $\log$(age) (log(Gyr))& -3 & 1.14 \\
        $\tau$ & -5 & 5 \\\hline
        Double power law SFH & & \\\hline
        $\log(\tau)$ & -4 & 2 \\
        A & -3 & 7 \\
        B & 0 & 7 \\
        C & 0 & 7 \\ \hline
        Piecewise SFH & & \\\hline
        $\log$(SFR) (log($M_\odot$/yr)) & -5 & 3.5 \\\hline
    \end{tabular}
    \caption{ MCSED priors used for the fitting here. We are assuming a uniform prior over the above specified bounds.}
    \label{tab:priors}
\end{table}

Since the walkers are initialized randomly across the entire allowed parameter space, the chains have significant burn-in. Based on visual inspection of the trace plots, a conservative estimate of the first 5000 steps is used as the burn-in period. In contrast to initializing the walkers in the region of maximum likelihood, the randomly chosen initial values ensure our walkers cover the entire parameter space and do not get caught in local maxima. As a trade-off, we must run our chains longer to accommodate the burn-in phase. 

To ensure that our parameter space is well sampled and that our fits have converged to the best values, we analyze the autocorrelation of our Markov chains. Since successive steps of Markov chains depend on the previous values, the steps are not independent. The autocorrelation time gives the length scale for which successive steps are not independent. Our effective number of samples is $N$ divided by the autocorrelation time. 

Following the recommendations of \citet{emcee}, we aim for each walker to sample for at least 10 autocorrelation times. The typical autocorrelation time for a given walker and parameter is approximately 550-600. This implies that our effective sample size is $N/600$ where $N$ is the number of samples. Therefore after removing the first 5000 samples of each walker for burn-in, the remaining 15000 samples equate to 25 autocorrelation times for each of our 500 walkers. This means we have $>12000$ independent samples when combining the different walkers. 

An example of a corner plot, which shows 2D projections of the high dimensional space, is displayed in Figure~\ref{fig:triangle}. The posterior distributions are constructed after removing the burn-in phase and are used for science.

\subsection{AGN activity}

\texttt{MCSED} does not model an AGN component. Other SED fitting codes handle potential AGN emission in different ways. For example, Prospector provides a procedure for modeling an obscured AGN via dust emission in the mid-IR \citep{2017ApJ...837..170L, 2021ApJS..254...22J}. However, since it does not model non-thermal X-ray/UV emission, it cannot model unobscured AGN. \citet{2020MNRAS.491..740Y} expanded CIGALE to fit unobscured AGN from the X-ray through IR. Since this work does not model AGN activity, we must take care that AGN contamination does not skew our results. \citet{2018ApJ...854...62L} investigated the effects of SED fitting obscured AGN and found that without an AGN component, various galaxy parameters were biased. 

\citet{2018ApJ...854...62L} used a sample of nearby galaxies from \citet{2014ApJS..212...18B}, of which there is some overlap with the SINGS/KINGFISH sample. From the overlap, \citet{2018ApJ...854...62L} identified 7 galaxies as having significant AGN contributions to the mid-IR spectra (NGC 2798, NGC 4254, NGC 4450, NGC 4569, NGC 4579, NGC 5033, and Mrk 33). To determine if there are significant unobscured AGN in our sample, we use the concurrent {\it Swift}/XRT observations to look for nuclear X-ray point sources \citep{2005SSRv..120..165B}. Using the 2SXPS catalog \citep{2020ApJS..247...54E}, we make a X-ray luminosity cut at $L_{X, \text{ 0.3-10 keV}}= 10^{42}$ erg/s to remove the brightest X-ray sources (NGC 1316, NGC 1404). For the galaxies without an X-ray detection nor overlap with the \citet{2014ApJS..212...18B} sample, we survey the literature to look for AGN classifications. This check flags two additional galaxies as probable AGN (NGC 1377 and NGC 3621) \citep{2009ApJ...700.1759G, 2012A&A...546A..68A}. In total, 11 galaxies are flagged as probable AGN with potentially significant contamination.

\citet{2010ApJS..190..233M} examined optical spectroscopy and classified the SINGS sample using emission line ratios \citep{1981PASP...93....5B}. By that classification, 35 galaxies were either AGN or composite objects. However, the SINGS/KINGFISH sample was constructed such that AGN activity contributes negligibly to the integrated spectrum of the galaxy \citep{2010ApJS..190..233M}. While some galaxies classified as having Low Ionization Nuclear Emission Regions (LINERs) potentially remain in this sample, our classification removes the most egregious interlopers. In summary, the following galaxies are removed from our SED fitting analysis due to suspected AGN activity: NGC 1316, NGC 1377, NGC 1404, NGC 2798, NGC 3621, NGC 4254, NGC 4450, NGC 4569, NGC 4579, NGC 5033, and Mrk 33. They are marked with a \dag\ symbol in Table~\ref{tab:fluxes}. 

For galaxies with significant AGN contributions to the SED, the integrated photometry can be fit using a multi-component model (stars, dust, gas, and AGN) in order to remove the AGN emission as done in works such as \citet{2013A&A...551A.100B, 2014MNRAS.439.2736D, 2017ApJS..233...19C, 2017MNRAS.470.4974D, 2018ApJ...857...64B}, among others. Due to the construction of the sample, we do not expect significant AGN contamination of the integrated light photometry.

Many previous works on this sample of galaxies have assumed that the integrated light is dominated by star formation and did not account for an AGN component \citep{Noll_09, 2011ApJ...738..124L, Hunt_SINGS_sed, 2021ApJ...920...96L}. For resolved scale studies, the central regions can be masked out to avoid the effects of AGN emission. Lastly, our main results do not change if we include these potential AGN in our analysis.

\section{Results} \label{sec:results}

We aim to characterize the dust attenuation curves of our sample of galaxies while minimizing the potential impact of the assumed form of the SFH on our conclusions. To this end, we fit each galaxy with each of the three SFHs described in \S\ref{sec:sfh}, with all other fitting assumptions held the same across the three fits.

Our analysis is done in multiple ways. First, we compare individual SED fits of a galaxy under different fitting assumptions and compare and contrast the results with the literature. Then, we investigate trends at the sample level and differences that may arise due to the different fitting assumptions. Next, we generate synthetic photometry and fit the mock data to test the ability of fits with different SFHs to accurately and precisely recover the attenuation law and other parameters like stellar mass. Lastly, we quantify the utility of the UVOT photometry by comparing fits with this additional data to fits without.

\subsection{Individual fits}

Many studies have looked at the dust attenuation law in nearby galaxies on semi-resolved scales: NGC 628 (M74), NGC 3031 (M81), NGC 3351 (M95), NGC 3034 (M82) \citep{Decleir2019, hoversten, 2021ApJ...913...37C, 2014MNRAS.440..150H}. We compare our results using integrated light photometry to these works to put our results in context with other results in the literature. However, differences in SED fitting, mainly the chosen SFH, and whether fitting on a global or resolved scale means that they are not necessarily directly comparable. 

Additionally, while a modified starburst law is widely used, the exact formulation can differ (see \S\ref{atten}). Here, we first add the bump to the starburst law and then modify by a power law. The current version of CIGALE \citep{2019A&A...622A.103B} modifies the starburst law first and then adds the bump. We use $E_b$ to describe the amplitude of the bump as parameterized in Equation~\ref{eq:noll_law} and $B$ for results using the current CIGALE implementation. 

Below, we quote the range in values measured using the three different SFHs. Table~\ref{tab:dust_param} lists the median of the posterior and standard deviation for $\delta$ and $E_b$ for each SFH.

NGC 628 is a face-on grand design spiral galaxy. \citet{Decleir2019} did pixel-by-pixel as well as global (i.e. integrated flux) SED fitting using CIGALE and found a global attenuation law with $\delta=-0.55 \pm 0.15$ and $B=2.24 \pm 1.16$. Here, we find $\delta$ between -0.08 and -0.33 and $E_b$ between 2.27 and 2.89. Our SFHs do not match as \citet{Decleir2019} used a delayed exponential SFH with added parameters to handle recent quenching or bursts of star formation. However, our derived slopes are shallower with our exponential SFH leading to a much shallower attenuation curve. 

NGC 3031 (M81), also known as Bode's galaxy, is a slightly inclined grand design spiral. \citet{hoversten} finds M81 is best fit with a MW-like dust law, which in the \citet{Noll_09} formalism corresponds to $\delta\sim-0.1$ and $E_b\sim3$ \citep{2020ARA&A..58..529S}. We find the attenuation curve is steeper than the MW law ($\delta$ between -0.28 and -0.37) with a small bump ($E_b$ between 0.61 and 2.94). There is good agreement on the slope fo the attenuation law between fits with different SFHs, with the piecewise SFH preferring the steepest attenuation curve with the strongest bump. One thing to note is that in \citet{hoversten}, they fit different static dust models (MW, SMC, LMC, Starburst) with varying $A_V$ rather than something like the \citet{Noll_09} dust law parameterization with a continuous variable describing the slope and bump. 

NGC 3351 is a barred spiral galaxy with a nuclear star-forming ring. \citet{2021ApJ...913...37C} investigated the central and nuclear regions and found an attenuation curve whose slope is similar to the MW. They do not fit for the dust bump due to UV spectroscopy that shows the bump to be insignificant. Here, our measurement agrees with the results from \citet{2021ApJ...913...37C} with $\delta$ between -0.13 and -0.26 and $E_b$ between 0.66 and 1.73. We note that within one standard deviation, all three SFH lead to attenuation laws consistent with no bump.

NGC 3034 (M82) is a starburst galaxy with the burst of star formation thought to be due to interactions with NGC 3031.  \citet{2014MNRAS.440..150H} found evidence for a MW-like 2175 \AA\ bump using color-color diagrams rather than SED fitting. As M82 is a starburst galaxy, the existence of a bump is contrary to the canonical starburst law. We find the attenuation curve is at least as steep as the MW and possible steeper than the SMC ($\delta$ between -0.11 and -0.64) with a MW-like bump ($E_b$ between 2.94 and 3.43). 

The broad agreement between the individual fits and the literature verifies that we are converging to the most probable SED model. The comparison to the literature is not one-to-one though; there will be differences between SED fits of integrated light versus pixel by pixel fitting and differences emerging from the underlying assumptions like SFH and the exact form of the attenuation law. 

\startlongtable
\begin{deluxetable*}{c|ccc|ccc}

\tablecaption{Dust law parameters $\delta$ and $E_b$ for each SFH \label{tab:dust_param}}
\tabletypesize{\scriptsize}
\tablehead{\\ Galaxy Name & \nocolhead{} & $\delta$ &  \nocolhead{} &  \nocolhead{} & $E_b$ & \nocolhead{} }
\startdata 
 & Exponential & Double power law & Piecewise & Exponential & Double power law & Piecewise \\\hline
  DDO 53 &  0.28 $\pm$ 0.38 &  0.32 $\pm$ 0.27 &  0.37 $\pm$ 0.23 & 0.87 $\pm$ 1.58 & 1.20 $\pm$ 1.58 & 1.04 $\pm$ 1.53 \\
 DDO 154 & -0.19 $\pm$ 0.33 & -0.19 $\pm$ 0.24 & -0.34 $\pm$ 0.44 & 2.33 $\pm$ 1.84 & 2.29 $\pm$ 1.80 & 2.65 $\pm$ 1.96 \\
 DDO 165 & -0.08 $\pm$ 0.35 & -0.07 $\pm$ 0.18 & -0.27 $\pm$ 0.30 & 0.60 $\pm$ 1.35 & 0.36 $\pm$ 0.93 & 1.07 $\pm$ 1.58 \\
 IC 2574 & -0.22 $\pm$ 0.26 & -0.25 $\pm$ 0.13 & -0.23 $\pm$ 0.23 & 2.98 $\pm$ 1.53 & 3.30 $\pm$ 1.34 & 3.12 $\pm$ 1.58 \\
 IC 4710 & -0.31 $\pm$ 0.65 &  0.09 $\pm$ 0.47 &  0.14 $\pm$ 0.35 & 2.25 $\pm$ 2.02 & 3.21 $\pm$ 1.88 & 3.44 $\pm$ 1.77 \\
    M81 dwA &  0.02 $\pm$ 0.37 &  0.05 $\pm$ 0.26 &  0.01 $\pm$ 0.42 & 2.65 $\pm$ 1.72 & 2.76 $\pm$ 1.64 & 2.45 $\pm$ 1.93 \\
   M101 & -0.09 $\pm$ 0.49 &  0.18 $\pm$ 0.21 &  0.02 $\pm$ 0.27 & 2.27 $\pm$ 1.91 & 1.19 $\pm$ 1.61 & 1.56 $\pm$ 1.45 \\
  NGC 24 & -0.19 $\pm$ 0.44 & -0.49 $\pm$ 0.39 & -0.14 $\pm$ 0.25 & 2.82 $\pm$ 1.82 & 2.91 $\pm$ 1.78 & 2.23 $\pm$ 1.65 \\
 NGC 337 & -0.59 $\pm$ 0.49 & -0.61 $\pm$ 0.30 & -0.50 $\pm$ 0.33 & 0.63 $\pm$ 1.58 & 1.13 $\pm$ 1.38 & 0.67 $\pm$ 1.35 \\
 NGC 584 & -0.50 $\pm$ 0.20 & -0.53 $\pm$ 0.12 & -0.58 $\pm$ 0.59 & 3.80 $\pm$ 1.23 & 4.16 $\pm$ 1.15 & 3.99 $\pm$ 1.58 \\
 NGC 628 & -0.08 $\pm$ 0.52 & -0.33 $\pm$ 0.35 & -0.20 $\pm$ 0.36 & 2.27 $\pm$ 1.97 & 2.89 $\pm$ 1.78 & 2.40 $\pm$ 1.54 \\
 NGC 855 & -0.80 $\pm$ 0.46 & -0.72 $\pm$ 0.40 & -0.81 $\pm$ 0.34 & 1.69 $\pm$ 1.72 & 2.25 $\pm$ 1.66 & 2.47 $\pm$ 1.62 \\
 NGC 925 & -0.22 $\pm$ 0.50 & -0.09 $\pm$ 0.43 & -0.16 $\pm$ 0.31 & 2.58 $\pm$ 1.95 & 2.53 $\pm$ 1.89 & 3.11 $\pm$ 1.68 \\
NGC 1097 & -0.05 $\pm$ 0.24 & -0.01 $\pm$ 0.10 & -0.12 $\pm$ 0.32 & 2.27 $\pm$ 1.28 & 2.17 $\pm$ 1.05 & 2.83 $\pm$ 1.53 \\
NGC 1266 & -0.55 $\pm$ 0.24 & -0.73 $\pm$ 0.20 & -0.77 $\pm$ 0.29 & 1.20 $\pm$ 1.83 & 0.42 $\pm$ 1.44 & 0.97 $\pm$ 1.74 \\
NGC 1291 & -0.30 $\pm$ 0.21 & -0.33 $\pm$ 0.09 & -0.65 $\pm$ 0.38 & 3.17 $\pm$ 1.08 & 3.50 $\pm$ 0.89 & 3.25 $\pm$ 1.79 \\
NGC 1512 &  0.00 $\pm$ 0.47 & -0.03 $\pm$ 0.18 &  0.03 $\pm$ 0.20 & 1.96 $\pm$ 2.01 & 2.68 $\pm$ 1.51 & 2.08 $\pm$ 1.32 \\
NGC 1566 & -0.09 $\pm$ 0.53 & -0.00 $\pm$ 0.29 & -0.06 $\pm$ 0.45 & 2.22 $\pm$ 1.97 & 2.62 $\pm$ 1.70 & 2.38 $\pm$ 1.49 \\
NGC 1705 & -0.24 $\pm$ 0.54 &  0.12 $\pm$ 0.18 &  0.11 $\pm$ 0.36 & 2.27 $\pm$ 2.00 & 1.53 $\pm$ 1.30 & 2.28 $\pm$ 1.95 \\
NGC 2146 & -0.17 $\pm$ 0.41 & -0.55 $\pm$ 0.32 & -0.44 $\pm$ 0.40 & 2.00 $\pm$ 1.89 & 1.90 $\pm$ 1.81 & 1.73 $\pm$ 1.71 \\
NGC 2403 & -0.43 $\pm$ 0.53 & -0.43 $\pm$ 0.27 & -0.34 $\pm$ 0.23 & 2.92 $\pm$ 1.99 & 3.71 $\pm$ 1.57 & 3.48 $\pm$ 1.47 \\
NGC 2841 & -0.21 $\pm$ 0.23 & -0.21 $\pm$ 0.09 & -0.27 $\pm$ 0.24 & 1.52 $\pm$ 1.00 & 1.37 $\pm$ 0.84 & 1.82 $\pm$ 1.60 \\
NGC 2915 & -0.07 $\pm$ 0.30 & -0.04 $\pm$ 0.12 &  0.00 $\pm$ 0.41 & 2.47 $\pm$ 1.34 & 2.64 $\pm$ 1.13 & 2.54 $\pm$ 1.95 \\
NGC 2976 & -0.57 $\pm$ 0.46 & -0.50 $\pm$ 0.34 & -0.21 $\pm$ 0.25 & 2.44 $\pm$ 1.83 & 2.87 $\pm$ 1.78 & 2.16 $\pm$ 1.47 \\
NGC 3031 & -0.28 $\pm$ 0.23 & -0.29 $\pm$ 0.10 & -0.37 $\pm$ 0.37 & 0.89 $\pm$ 1.27 & 0.61 $\pm$ 1.04 & 2.75 $\pm$ 1.90 \\
NGC 3034 & -0.50 $\pm$ 0.46 & -0.64 $\pm$ 0.25 & -0.11 $\pm$ 0.39 & 3.20 $\pm$ 1.57 & 3.43 $\pm$ 1.40 & 2.94 $\pm$ 1.45 \\
NGC 3049 & -0.46 $\pm$ 0.49 & -0.38 $\pm$ 0.36 & -0.29 $\pm$ 0.29 & 1.46 $\pm$ 1.81 & 1.41 $\pm$ 1.66 & 1.50 $\pm$ 1.65 \\
NGC 3184 & -0.24 $\pm$ 0.58 & -0.04 $\pm$ 0.30 &  0.03 $\pm$ 0.37 & 2.24 $\pm$ 2.05 & 2.40 $\pm$ 1.84 & 1.22 $\pm$ 1.37 \\
NGC 3190 & -0.33 $\pm$ 0.25 & -0.49 $\pm$ 0.17 & -0.40 $\pm$ 0.30 & 3.26 $\pm$ 1.30 & 3.81 $\pm$ 1.28 & 3.31 $\pm$ 1.66 \\
NGC 3198 & -0.12 $\pm$ 0.57 & -0.04 $\pm$ 0.30 &  0.09 $\pm$ 0.24 & 1.93 $\pm$ 1.98 & 3.00 $\pm$ 1.74 & 2.05 $\pm$ 1.54 \\
NGC 3265 & -0.65 $\pm$ 0.43 & -0.37 $\pm$ 0.22 & -0.38 $\pm$ 0.27 & 2.04 $\pm$ 1.61 & 1.91 $\pm$ 1.33 & 1.86 $\pm$ 1.32 \\
NGC 3351 & -0.13 $\pm$ 0.51 & -0.25 $\pm$ 0.18 & -0.26 $\pm$ 0.28 & 1.73 $\pm$ 2.06 & 0.66 $\pm$ 1.38 & 1.02 $\pm$ 1.47 \\
NGC 3521 & -0.14 $\pm$ 0.23 & -0.36 $\pm$ 0.18 & -0.14 $\pm$ 0.30 & 0.83 $\pm$ 1.23 & 1.94 $\pm$ 1.34 & 1.59 $\pm$ 1.49 \\
NGC 3627 & -0.27 $\pm$ 0.36 & -0.41 $\pm$ 0.23 & -0.16 $\pm$ 0.29 & 0.79 $\pm$ 1.42 & 1.62 $\pm$ 1.37 & 1.16 $\pm$ 1.19 \\
NGC 3773 & -0.21 $\pm$ 0.51 & -0.17 $\pm$ 0.35 & -0.20 $\pm$ 0.37 & 2.07 $\pm$ 1.94 & 1.66 $\pm$ 1.86 & 2.13 $\pm$ 1.86 \\
NGC 3938 & -0.23 $\pm$ 0.53 & -0.14 $\pm$ 0.39 & -0.07 $\pm$ 0.39 & 2.23 $\pm$ 2.01 & 1.36 $\pm$ 1.98 & 0.45 $\pm$ 1.42 \\
NGC 4125 & -0.75 $\pm$ 0.25 & -0.87 $\pm$ 0.17 & -0.60 $\pm$ 0.62 & 0.77 $\pm$ 1.40 & 0.30 $\pm$ 1.10 & 1.86 $\pm$ 2.06 \\
NGC 4236 &  0.00 $\pm$ 0.29 & -0.04 $\pm$ 0.17 & -0.09 $\pm$ 0.30 & 1.42 $\pm$ 1.37 & 1.61 $\pm$ 1.35 & 2.10 $\pm$ 1.72 \\
NGC 4321 &  0.15 $\pm$ 0.39 & -0.18 $\pm$ 0.22 & -0.16 $\pm$ 0.32 & 0.68 $\pm$ 1.73 & 0.73 $\pm$ 1.34 & 1.14 $\pm$ 1.65 \\
NGC 4536 &  0.12 $\pm$ 0.41 & -0.06 $\pm$ 0.20 & -0.07 $\pm$ 0.32 & 1.58 $\pm$ 1.81 & 2.10 $\pm$ 1.62 & 1.89 $\pm$ 1.58 \\
NGC 4552 & -0.40 $\pm$ 0.27 & -0.41 $\pm$ 0.12 & -0.30 $\pm$ 0.50 & 4.29 $\pm$ 1.50 & 4.77 $\pm$ 1.02 & 3.08 $\pm$ 2.11 \\
NGC 4559 & -0.14 $\pm$ 0.50 &  0.03 $\pm$ 0.32 &  0.09 $\pm$ 0.17 & 2.34 $\pm$ 1.99 & 1.53 $\pm$ 1.85 & 1.04 $\pm$ 1.31 \\
NGC 4594 & -0.43 $\pm$ 0.15 & -0.45 $\pm$ 0.10 & -0.66 $\pm$ 0.40 & 3.65 $\pm$ 1.39 & 4.00 $\pm$ 1.23 & 2.61 $\pm$ 1.93 \\
NGC 4625 & -0.43 $\pm$ 0.47 & -0.03 $\pm$ 0.42 & -0.11 $\pm$ 0.44 & 3.00 $\pm$ 1.96 & 1.83 $\pm$ 1.98 & 2.43 $\pm$ 1.95 \\
NGC 4631 &  0.17 $\pm$ 0.41 &  0.21 $\pm$ 0.21 &  0.15 $\pm$ 0.27 & 1.37 $\pm$ 1.65 & 1.20 $\pm$ 1.35 & 1.15 $\pm$ 1.33 \\
NGC 4725 & -0.14 $\pm$ 0.26 & -0.12 $\pm$ 0.08 & -0.21 $\pm$ 0.20 & 1.08 $\pm$ 1.00 & 0.97 $\pm$ 0.83 & 2.04 $\pm$ 1.49 \\
NGC 4736 & -0.08 $\pm$ 0.50 & -0.34 $\pm$ 0.20 & -0.24 $\pm$ 0.36 & 2.20 $\pm$ 1.96 & 3.11 $\pm$ 1.81 & 2.48 $\pm$ 1.72 \\
NGC 4826 & -0.42 $\pm$ 0.41 & -0.46 $\pm$ 0.24 & -0.30 $\pm$ 0.24 & 1.42 $\pm$ 1.49 & 2.54 $\pm$ 1.26 & 2.57 $\pm$ 1.33 \\
NGC 5055 &  0.11 $\pm$ 0.28 & -0.18 $\pm$ 0.14 & -0.14 $\pm$ 0.26 & 0.90 $\pm$ 1.43 & 2.08 $\pm$ 1.40 & 1.32 $\pm$ 1.35 \\
NGC 5194 &  0.01 $\pm$ 0.50 & -0.27 $\pm$ 0.28 & -0.29 $\pm$ 0.49 & 1.78 $\pm$ 2.16 & 1.79 $\pm$ 1.67 & 1.40 $\pm$ 1.45 \\
NGC 5398 & -0.04 $\pm$ 0.48 & -0.08 $\pm$ 0.48 & -0.15 $\pm$ 0.37 & 2.00 $\pm$ 1.92 & 2.04 $\pm$ 1.93 & 1.95 $\pm$ 1.82 \\
NGC 5474 &  0.01 $\pm$ 0.33 &  0.01 $\pm$ 0.25 & -0.24 $\pm$ 0.44 & 0.75 $\pm$ 1.37 & 0.99 $\pm$ 1.52 & 1.77 $\pm$ 1.61 \\
NGC 5713 & -0.29 $\pm$ 0.39 & -0.73 $\pm$ 0.26 & -0.14 $\pm$ 0.31 & 0.68 $\pm$ 1.50 & 2.36 $\pm$ 1.69 & 1.21 $\pm$ 1.49 \\
NGC 5866 & -0.50 $\pm$ 0.23 & -0.55 $\pm$ 0.14 & -0.88 $\pm$ 0.33 & 2.42 $\pm$ 0.94 & 2.63 $\pm$ 0.87 & 3.67 $\pm$ 1.37 \\
NGC 6822 & -0.21 $\pm$ 0.54 & -0.10 $\pm$ 0.49 & -0.09 $\pm$ 0.51 & 1.98 $\pm$ 1.92 & 1.81 $\pm$ 1.95 & 1.97 $\pm$ 1.95 \\
NGC 6946 & -0.44 $\pm$ 0.60 & -0.34 $\pm$ 0.36 & -0.20 $\pm$ 0.36 & 0.20 $\pm$ 1.28 & 0.20 $\pm$ 1.13 & 0.41 $\pm$ 1.20 \\
NGC 7331 & -0.29 $\pm$ 0.33 & -0.37 $\pm$ 0.21 & -0.17 $\pm$ 0.30 & 1.22 $\pm$ 1.54 & 1.80 $\pm$ 1.37 & 0.86 $\pm$ 1.06 \\
NGC 7552 & -0.05 $\pm$ 0.33 &  0.05 $\pm$ 0.20 & -0.15 $\pm$ 0.34 & 1.93 $\pm$ 1.34 & 1.79 $\pm$ 1.18 & 2.27 $\pm$ 1.35 \\
NGC 7793 &  0.09 $\pm$ 0.44 &  0.13 $\pm$ 0.22 &  0.25 $\pm$ 0.24 & 1.51 $\pm$ 1.79 & 1.65 $\pm$ 1.48 & 1.23 $\pm$ 1.43 \\
UGC 4305 & -0.22 $\pm$ 0.31 & -0.15 $\pm$ 0.16 & -0.22 $\pm$ 0.23 & 2.52 $\pm$ 1.59 & 2.59 $\pm$ 1.40 & 2.74 $\pm$ 1.67 \\
UGC 5139 & -0.02 $\pm$ 0.26 & -0.01 $\pm$ 0.16 & -0.08 $\pm$ 0.36 & 2.63 $\pm$ 1.51 & 2.64 $\pm$ 1.44 & 2.60 $\pm$ 1.89 \\
UGC 5336 & -0.20 $\pm$ 0.43 & -0.17 $\pm$ 0.35 & -0.07 $\pm$ 0.49 & 2.07 $\pm$ 1.87 & 2.23 $\pm$ 1.88 & 2.20 $\pm$ 2.02 \\
UGC 5423 & -0.05 $\pm$ 0.29 & -0.03 $\pm$ 0.12 & -0.15 $\pm$ 0.16 & 2.10 $\pm$ 1.11 & 2.17 $\pm$ 0.94 & 2.82 $\pm$ 1.31
\enddata
\end{deluxetable*}

\subsection{Sample level analysis}

\begin{figure*}
    \centering
    \includegraphics[width=0.9\textwidth]{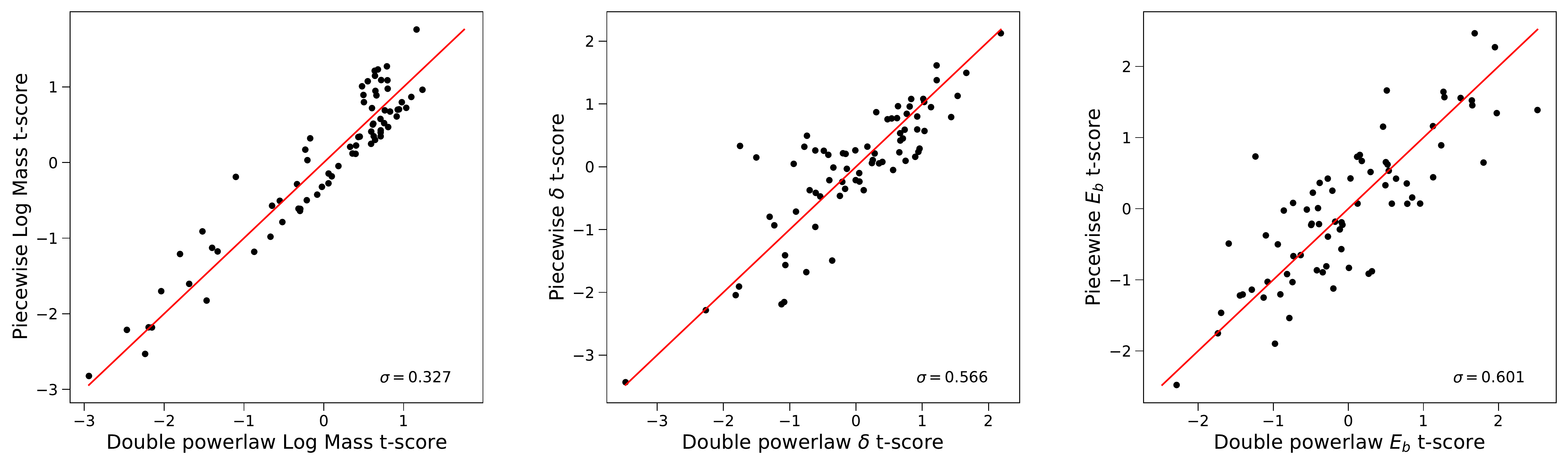}
    \caption{Comparison of the piecewise and double power law fits using standardized t-scores for stellar mass, $\delta$, and $E_b$. We find less scatter in stellar mass compared to the dust law parameters for every possible pair of SFHs. This means that parameters such as stellar mass are much less sensitive to the chosen SFH.} 
    \label{fig:disp_comp}
\end{figure*}

Next, we look at the individual posteriors for the dust law parameters ($E(B-V), \delta$, and $E_b$), stellar mass, and recent SFR to see how the choice of SFH affects these parameters. Large scatter or a bias when comparing fits using a different SFH would demonstrate which parameters are most sensitive to the form of the SFH. We standardize the best-fit values by subtracting the sample mean and dividing by the standard deviation. This leads to dimensionless numbers that we can then use to compare the scatter between parameters. In Figure~\ref{fig:disp_comp}, we show the comparison of the fits using the piecewise SFH and the double power law SFH for the parameters stellar mass, $\delta$, and $E_b$. When comparing the scatter between the fits for these parameters, we find a larger scatter for the dust law parameters than for stellar mass, implying that the stellar mass is more robust to the choice of SFH as compared to the dust curve parameters. This was true for the other pairs of SFHs as well.

\begin{figure*}
    \begin{tabular}{c c}
         \includegraphics[width=0.45\textwidth]{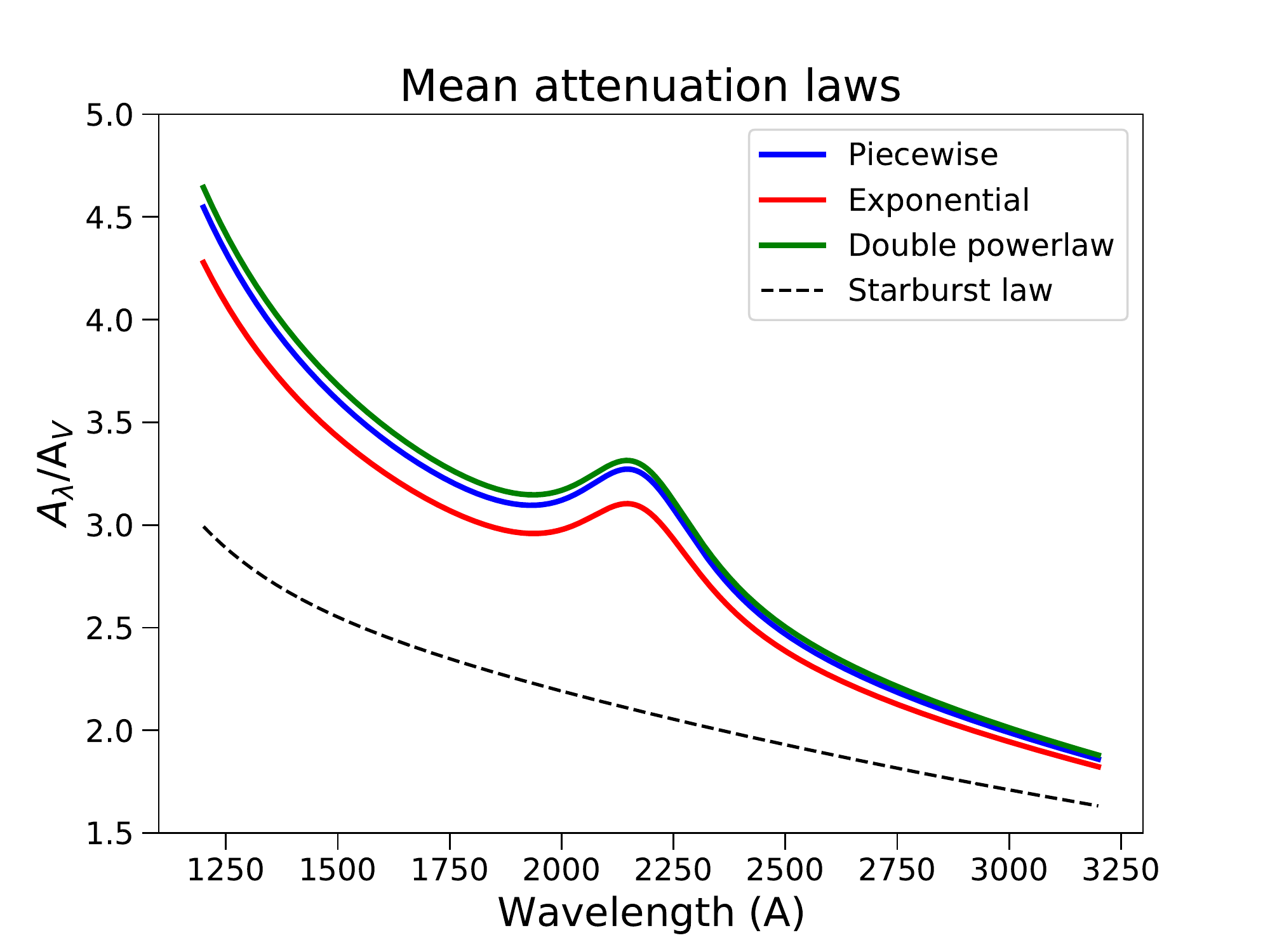} &         \includegraphics[width=0.45\textwidth]{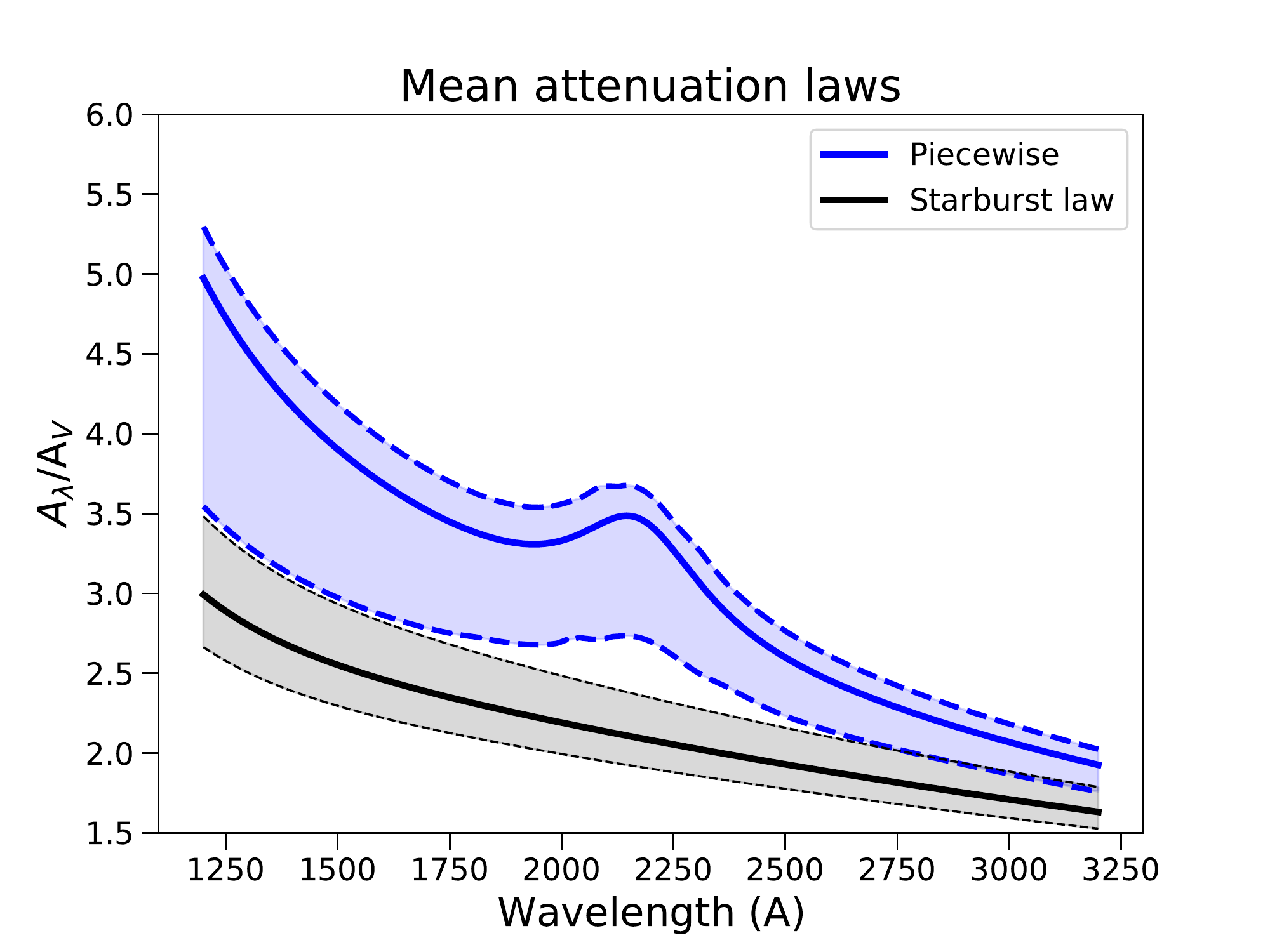} \\
     \end{tabular}
    \caption{Average attenuation curves for the SINGS/KINGFISH sample. {\it Left:} Mean attenuation law of entire sample, calculated by averaging over the entire sample at each wavelength for each different SFH tested. {\it Right:} Mean attenuation law from the piecewise SFH fits compared to the starburst law, excluding galaxies consistent with no dust. The 16th and 84th percentiles are shown for the piecewise SFH result and the \citet{C00} error in $R_V$ is displayed. We see that the derived attenuation law for each SFH is steeper than the \citet{C00} law.}
    \label{fig:average_laws}
\end{figure*}

From the posterior samples, we take the median of the posterior distribution for each galaxy to get $\delta$ and $E_b$. Using the \citet{Kriek+Conroy} parameterization of the dust law, we construct the attenuation curve ($A_{\lambda}/A_V$) of each galaxy and compute the sample mean for each of the three SFH fits at a 1 \AA\ resolution. The left panel of Figure~\ref{fig:average_laws} shows the mean attenuation law using each of the SFHs with the starburst law as a reference. Broadly, we see good agreement between the fits using different SFHs and note that the exponential SFH leads to a slightly shallower average attenuation law. Assuming $A_V=1$, fits using the different SFHs result in sample-averaged attenuation curves that vary by up to 0.25 mag at 1500 \AA.

We fit the average curve with the \citet{Noll_09} law to determine ($\delta$, $E_b$) and we find ($-0.25^{+0.31}_{-0.24}, 1.80^{+0.70}_{-0.38}$), ($-0.22^{+0.29}_{-0.19}, 1.80^{+0.53}_{-0.38}$), and ($-0.18^{+0.27}_{-0.23}, 1.64^{+0.62}_{-0.38}$) for the double power law, piecewise, and exponential SFHs respectively. The quoted errors are the 16th and 84th percentiles.

\citet{Decleir2019} found that constraints on the dust law drastically improved when only focused on regions with $A_V>0.2$, akin to removing dust free regions. This $A_V$ threshold corresponds to roughly $E(B-V)\sim0.05$. We have some galaxies in our sample that are consistent with no dust for at least one SFH. The exponential SFH fits had a larger fraction of galaxies consistent with no dust. To minimize the number of galaxies removed from our sample, we remove dust free galaxies by setting our threshold at $E(B-V)<0.03$. We then calculate the average attenuation law for the remaining 47 galaxies. Removing these dust free galaxies does not meaningfully change the result. The right panel in Figure~\ref{fig:average_laws} shows the average attenuation law from fits using the piecewise SFH after removing potentially dust free galaxies. The spread for the attenuation law from the piecewise SFh are the 16th and 84th percentiles and the uncertainty on $R_V$ from \citet{C00} is shown as well. With the dust free galaxies removed and again assuming $A_V=1$, different SFH choices lead to $A_{1500}$ varying by up to 0.2 mag. Comparing the average dust curve from the double power law fits, which led to the steepest attenuation curve, to the starburst law, we find the deviation at 1500 \AA\ is 1.36 mag.

In terms of ($\delta$, $E_b$) after removing dust-free galaxies, we find ($-0.31^{+0.31}_{-0.24}, 1.84^{+0.91}_{-0.33}$), ($-0.28^{+0.21}_{-0.18} , 1.91^{+0.51}_{-0.32}$), and ($-0.27^{+0.27}_{-0.23}, 1.67^{+0.64}_{-0.29}$) for the double power law, piecewise, and exponential SFHs respectively. The errors reported are based on the 16th and 84th percentiles.

The average attenuation law of the sample using different SFHs agree within errors. However, when there is a discrepancy, we see anecdotal evidence that the exponential SFH tends to prefer shallower slopes. To quantify this, we do a one sided t-test to determine if the difference $\delta_{EXP}-\delta_{DPL}$ has a mean greater than 0, corresponding to the exponential SFH leading to shallower slopes. We find significance with a p-value of 0.0087. While this test cannot determine which SFH is the most appropriate choice, it demonstrates that the choice of SFH in SED fitting can lead to systematic effects in derived properties. Future work on a larger sample will be needed.

Due to the limited sample size, we are limited in our ability to test relationships between the attenuation law slope and $A_V$, slope and bump strength, slope and inclination angle, or slope and metallicity \citep[as was e.g. done by][]{Salim_2018, Kriek+Conroy, 2017ApJ...851...90B, decleir_thesis, 2020ApJ...899..117S}. Nevertheless, we will provide such a study in future work (Decleir et al., in prep.).

\subsection{Parameter recovery using synthetic galaxies}

To test the ability of different SFHs to recover the true dust law parameters and other galaxy properties like stellar mass, we used \texttt{MCSED} to generate and fit mock photometry. Fitting synthetic photometry allows us to quantify how well we are able to recover known parameters. One caveat is that the photometry is generated using the idealized SFH parameterizations. In practice, we do not know the form of the SFH but must assume some form to approximate real complex epochs of star formation. \citet{lower20} investigated the effect that different SFHs had on stellar mass and found that standard parametric SFHs underestimated stellar mass compared to the flexible, piecewise SFH defined in \citet{2019ApJ...876....3L}. 

In this analysis, we generate synthetic photometry for 150 mock galaxies for each of the three different SFHs, for a total of 450 simulated galaxy SEDs. Using the median values for every galaxy determined from our fits to the real data, we calculate the sample mean and standard deviation for every parameter and draw the mock galaxy parameters from a normal distribution mirroring the observed values. Additionally, we use half as many walkers as described in \S\ref{sec:SED_fitting} since the fits converge faster due to the exact match between the parametric forms used in simulating and fitting the data. We fit the generated data using the same parameterization of the SFH as was used to generate the photometry, holding all other SED fitting options constant. We then quantify the precision and accuracy of recovering the dust law parameters when the SFH form is known to be correct. 

\begin{figure*}
    \centering
    \includegraphics[width=\textwidth]{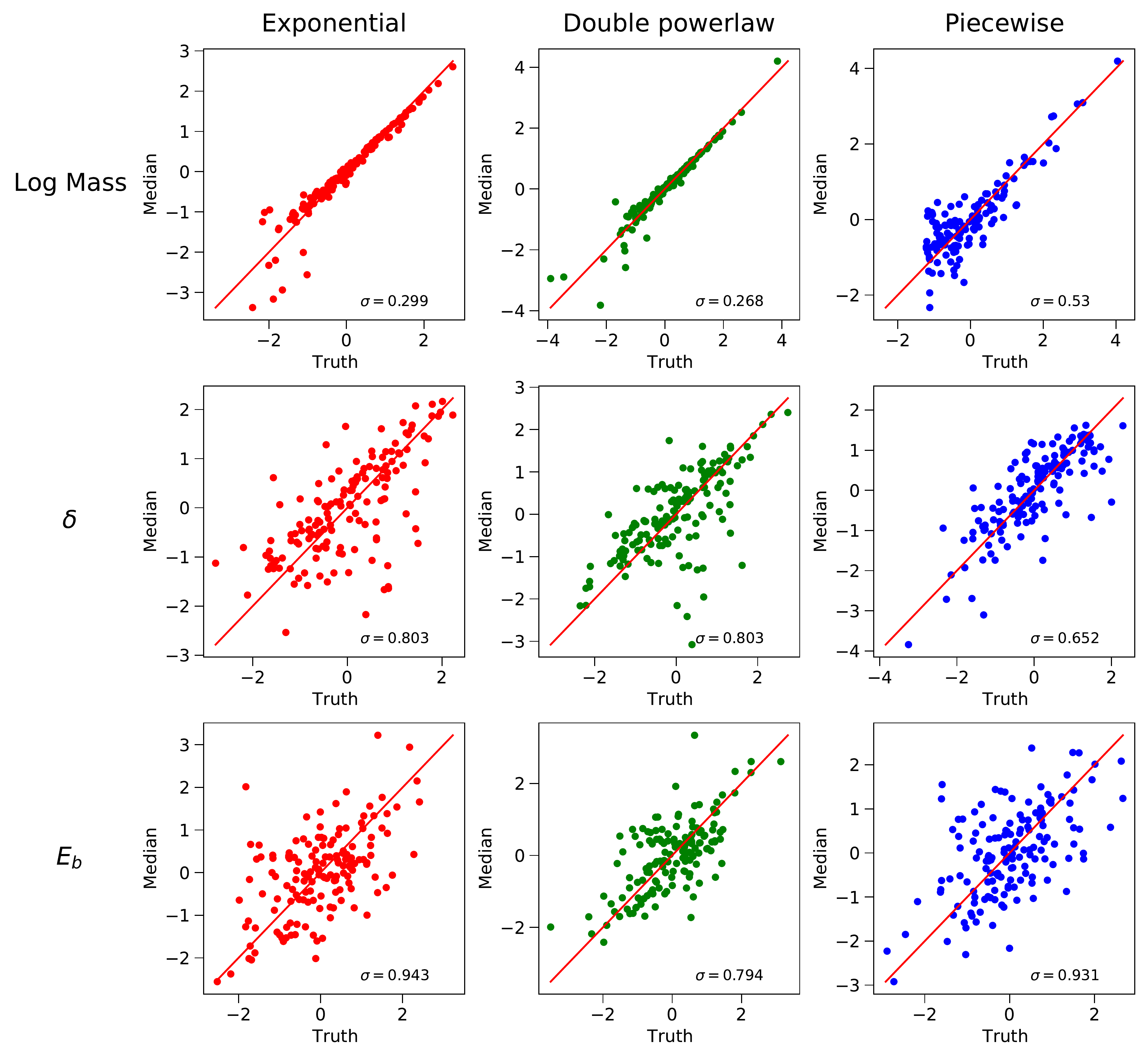}
    \caption{Matrix of plots showing the median fitted values vs the true values from synthetic galaxy fits. The values are standardized using the sample mean and standard deviation so that the scatter in stellar mass can be compared to the slope and bump strength scatter. Shown are the parameters stellar mass, $\delta$, and $E_b$ for each of the star formation history assumptions. We find that the exponential SFH, while being the simplest model, did not lead to the most precise parameter recovery for stellar mass, $\delta$, or $E_b$. Additionally, the most complex SFH, the piecewise, had the smallest scatter in recovering $\delta$.} 
    \label{fig:test_comp}
\end{figure*}

Figure~\ref{fig:test_comp} shows a comparison of the median fitted and true values for the stellar mass and dust law parameters for each of the different SFHs for mock galaxies. The fits using the double power law SFH showed the least scatter for the stellar mass and bump strength parameters. Interestingly, the piecewise SFH led to the best recovery of the slope of the attenuation curve. The exponential SFH fits, which had the fewest parameters, were the least precise in recovering the bump strength and slightly worse than the double power law SFH at recovering the stellar mass. This suggests that despite the exponential SFH being the simplest model, it leads to scatter in parameter recovery similar to or larger than more complex models. It is encouraging that regardless of the SFH, the derived galaxy parameters are not biased. 

Choosing an appropriate star formation history is important for robust measurements of galaxy physical properties as has been shown in \citet{2019ApJ...873...44C} and \citet{2019ApJ...876....3L} which discussed the pitfalls of parametric versus nonparametric SFHs and concluded that nonparametric SFHs like the piecewise form recover true SFHs more accurately and with less bias than typical parametric forms. Additionally, \citet{lower20} found that parametric SFHs were biased in recovering stellar mass compared to nonparametric models. As SFHs are expected to be complex, the most flexible parameterization is likely to give the best result with a computational trade off due to the extra parameters. For example, the double power law SFH assumes a peak of star formation and then a decline to present day levels. However, it cannot handle multiple intense bursts of star formation due to mergers or accretion of pristine gas, while the piecewise SFH would better be able to match that complex history. Based on previous results in the literature and the scatter seen in our piecewise fits, we find that the piecewise SFH is a viable choice, especially with the wealth of data for nearby objects balancing the additional parameters required. However, in these idealized tests, the double power law also performed well and would be a reasonable choice given the data and model used here.

More complex tests can be done such as having the true SFH form differ from the one used for fitting. \citet{2019ApJ...876....3L} showed that the piecewise SFH was able to match a variety of true underlying SFH shapes and showed less bias than other parametric SFHs. \citet{Decleir2019} did a similar test using two different parametric SFHs (one for creating the mock data and another one for the fitting) and found that there tended to be an increase in the scatter between fitted and true values, compared to using the same SFH for creating and fitting the mock data. Tests such as these highlight the importance of justifying ones assumptions when performing SED fitting.

\subsection{Impact of UVOT photometry}

Lastly, we quantify the utility of the extra UV data provided by {\it Swift}/UVOT.  We do this for both our sample of galaxies and simulated galaxy SEDs. We look if the inclusion of the UVOT photometry meaningfully changes the median value of the slope of the attenuation law or the bump strength and if the additional data increases precision in the parameter estimates. 

First, we look to see if there are large systematic shifts in the dust law parameters when comparing SED fits of our real data with and without the UVOT data. We find that for $\delta$, the power law deviation from the starburst law, the median shift is minimal ($\delta_{\text{no UVOT}} - \delta_{\text{UVOT}} \sim -0.05$) with a standard deviation of $\sim0.1$. This trend holds for all three SFHs tested. 

For the bump strength, we see a shift towards higher bump strengths when removing the UVOT photometry. On average for each SFH, the bump strength with UVOT is 2/3 to 3/4 the bump strength without UVOT data, where ($E_{b,\text{ UVOT}})/ (E_{b,\text{ no UVOT}}) \sim 0.65, 0.74, 0.77$ for the piecewise, double power law, and exponential SFHs respectively.

The additional UV photometry at and on either side of the 2175 \AA\ dust bump is expected to lead to tighter constraints on the bump strength.  We calculate the percent change in the uncertainty of $E_b$ when including the UVOT photometry. As expected, the median uncertainty in $E_b$ decreased by 12-16\% depending on the SFH. The exponential SFH showed the smallest decrease in scatter when including the UVOT data. 

\begin{figure*}
    \centering
    \includegraphics[width=0.9\textwidth]{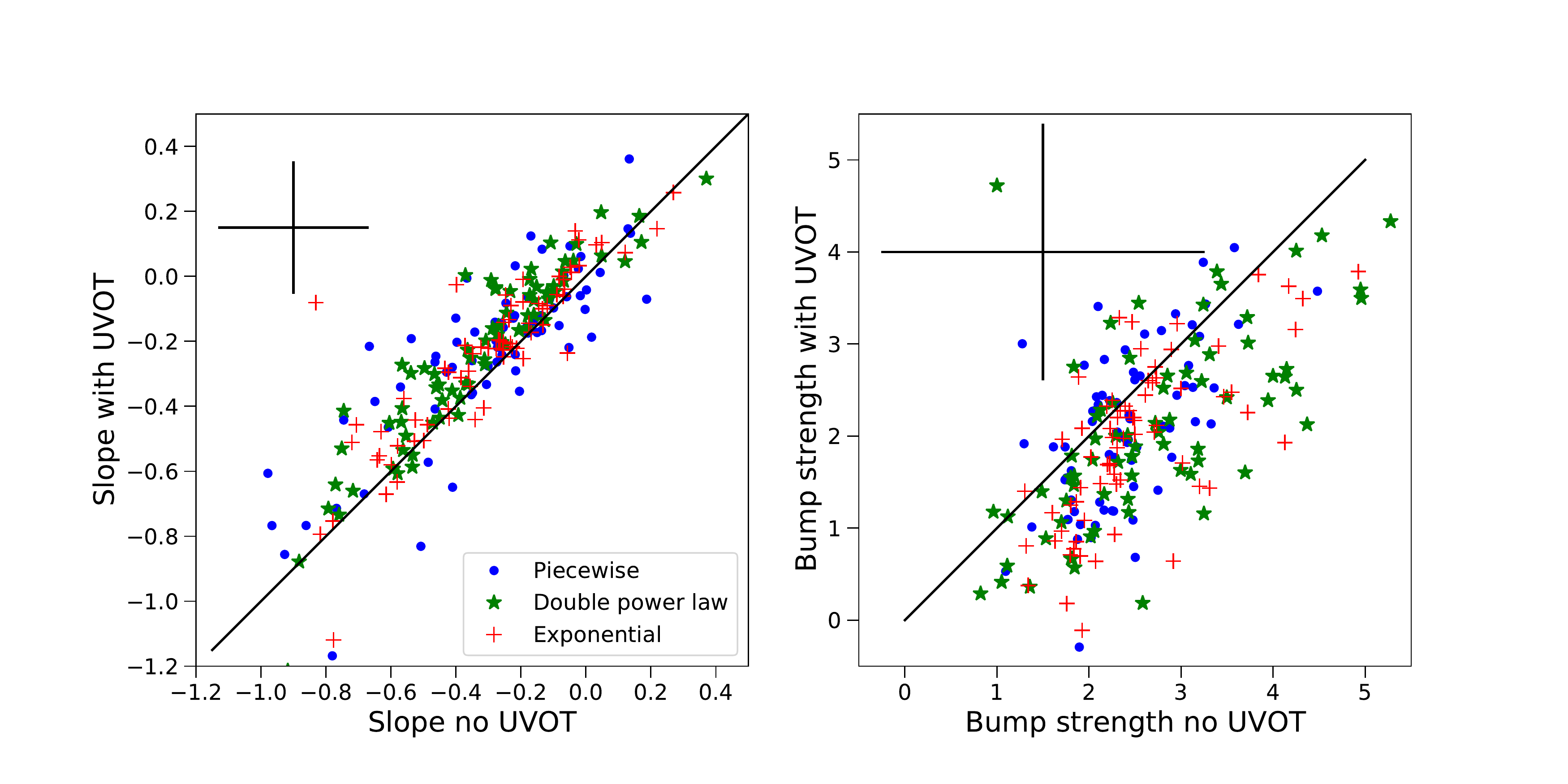}
    \caption{Comparison of dust law parameters when fit with and without UVOT data. {\it Left:} The attenuation law slope with UVOT vs the slope when fit without UVOT. {\it Right:} The 2175 $\AA\ $ bump strength with UVOT vs the bump from fits without UVOT data. We see that there is fairly good agreement on the slope of the attenuation law between fits with and without UVOT data. However, there is a noticeable difference when including the UVOT data for the bump strength. We find that fits with the UVOT data lead to bump strengths that are $\sim2/3$ to $3/4$ of the bump as determined without UVOT photometry. This could point towards the inability to accurately measure the bump strength without additional near-UV data. Characteristic errors are shown in the upper left of each plot. We see that including UVOT leads to a reduction in the size of the error bars.} 
    \label{fig:bump_offset}
\end{figure*}

The comparison of the dust law parameters for the sample with and without UVOT data and for each SFH are shown in Figure~\ref{fig:bump_offset}. Regardless of which SFH is used, we find that the bump strength is systematically higher in fits without the UVOT data compared to fits including UVOT. No such offset is seen with $\delta$. This highlights the strength of UVOT data in recovering the bump strength parameter. Due to the uniform prior assumed for SED fitting, it appears the fits with {\it GALEX}-alone cannot constrain the bump, leading to bump strengths tending towards the median of the priors ($E_b=2.5$).

\begin{figure}
    \centering
    \includegraphics[width=0.48\textwidth]{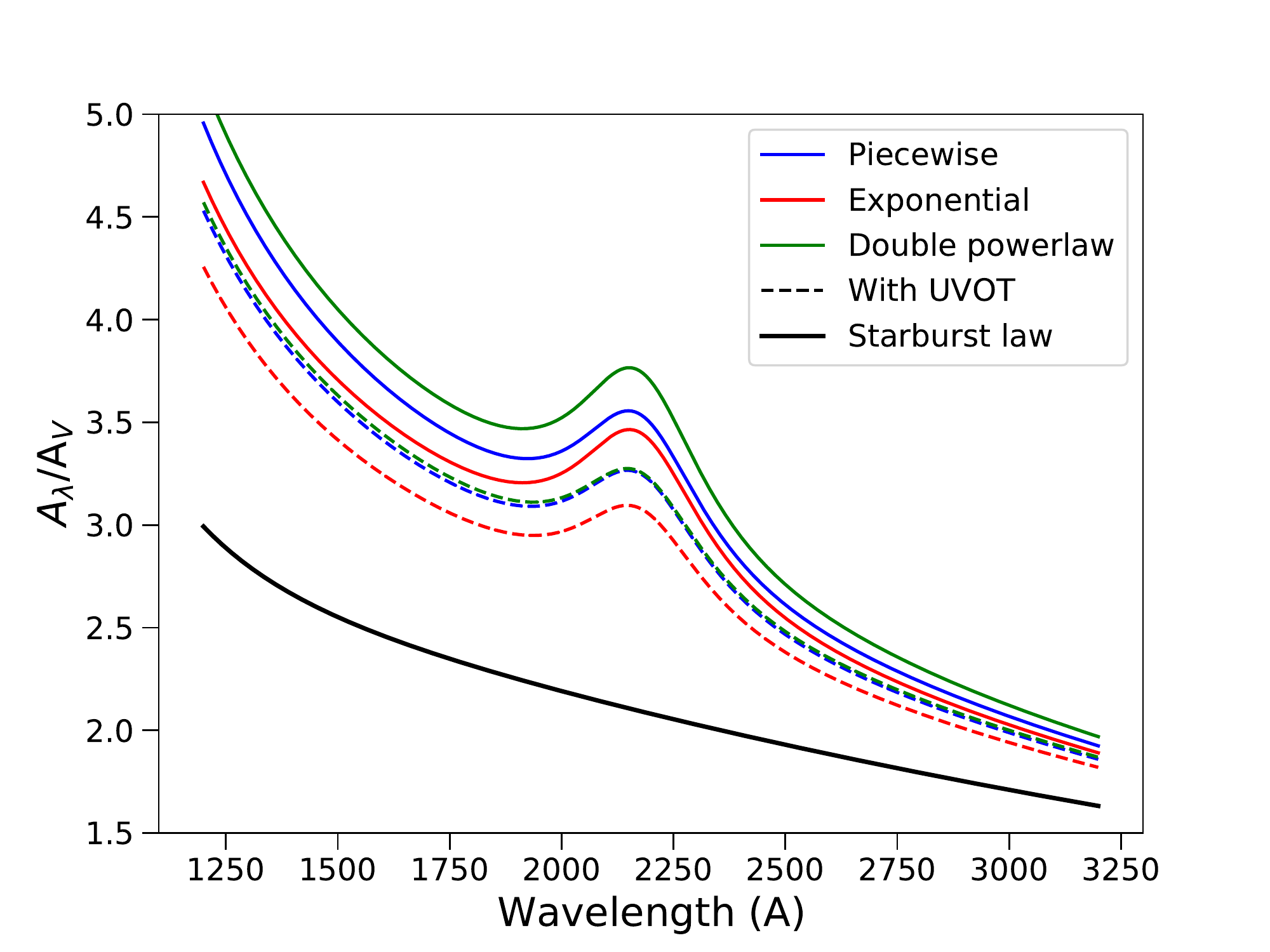}
    \caption{Sample average attenuation law compared from fits with and without UVOT photometry. We find that the average bump strength is smaller when including the UVOT data. We also see a small shift to shallower slopes. This means that when determining the attenuation curve without UVOT data, it is likely that one will overestimate the amount of attenuation at all wavelengths. This is compared to the starburst law, which significantly underestimates attenuation, regardless of photometry used and SFH parameterization assumed.}
    \label{fig:sample_wout}
\end{figure}

We also see differences in the sample average attenuation curve when fitting SEDs without the UVOT photometry. Figure~\ref{fig:sample_wout} shows the sample average attenuation law as derived from fits with and without the UVOT photometry. The fits without UVOT have a much larger bump strength and we still see a trend between the slope of the attenuation law with chosen SFH parameterization. There is also a marginal effect on the slope based on the inclusion of UVOT data. For the fits without UVOT data, the difference between the attenuation laws at 1500 \AA\ from the piecewise and exponential fits, assuming $A_V=1$, is 0.41 mag. 

For the average attenuation laws shown in Figure~\ref{fig:sample_wout}, the difference between the steepest slope from fits without UVOT data and the shallowest from fits with UVOT data in $A_{\lambda}/A_V$ is 0.64 mag at 1500 $\AA$ for $A_V=1$. This means that both the data included during SED fitting and the assumptions made can lead to a substantial difference in the amount of attenuation expected at $1500$ \AA. When the steepest average law from fits without UVOT data is compared to the starburst law, the difference is 1.5 mag at 1500 \AA for $A_V=1$. 

\begin{figure}
    \centering
    \includegraphics[width=0.5\textwidth]{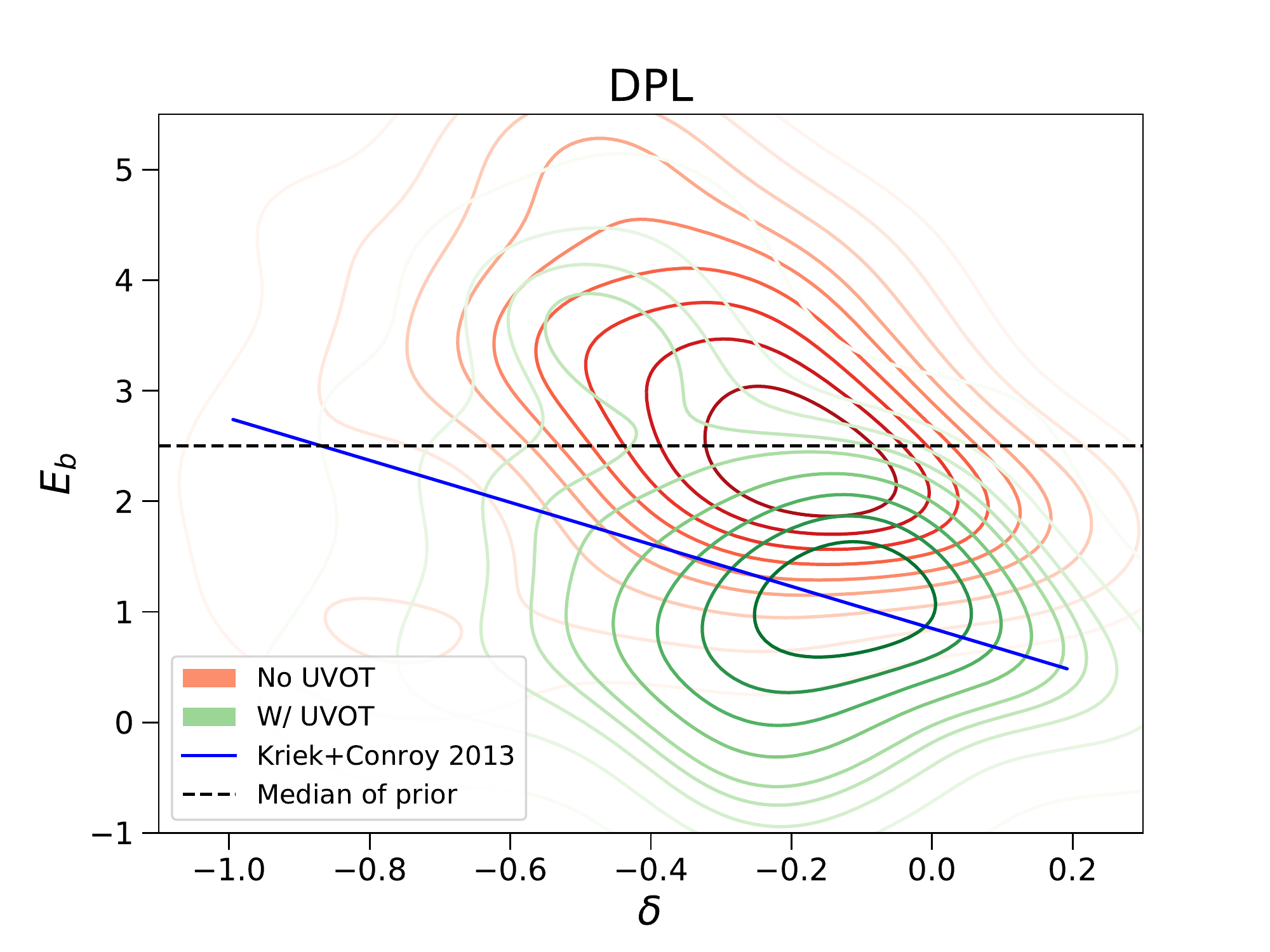}
    \caption{Density plots in the $\delta-E_b$ plane for the fits using the double power law SFH. The values from fits including UVOT data are in line with the expectation from \citet{Kriek+Conroy}, while the fits without lie systematically above. The black dashed line shows the median of the uniform bound prior. The bump strengths from the fits without UVOT data are clustered around that point, implying that the fits without UVOT data cannot constrain the bump beyond the uninformative prior.}
    \label{fig:dpl_KC13}
\end{figure}

We are further convinced of the importance of UVOT data after comparing our derived dust law parameters to the trend between the slope and bump strength seen in \citet{Kriek+Conroy}. Figure~\ref{fig:dpl_KC13} contains density plots in the $\delta-E_b$ plane for the double power law SFH, which is the SFH that showed the least scatter. The trend seen in the fits including UVOT are a better match to the \citet{Kriek+Conroy} relation than the fits without. However, other studies like \citet{2016ApJ...833..201S} have found different relationships in the $\delta-E_b$ plane. What we do see is that the bump strength from fits without UVOT is consistent with the middle of the range allowed by the uniform prior, implying that the data cannot constrain the bump beyond the uninformative prior. This trend where the bump strength as determined from fits without UVOT photometry was distributed around the median of the prior is consistent for each of the three SFHs tested here.

\begin{figure*}
    \centering
    \includegraphics[width=0.9\textwidth]{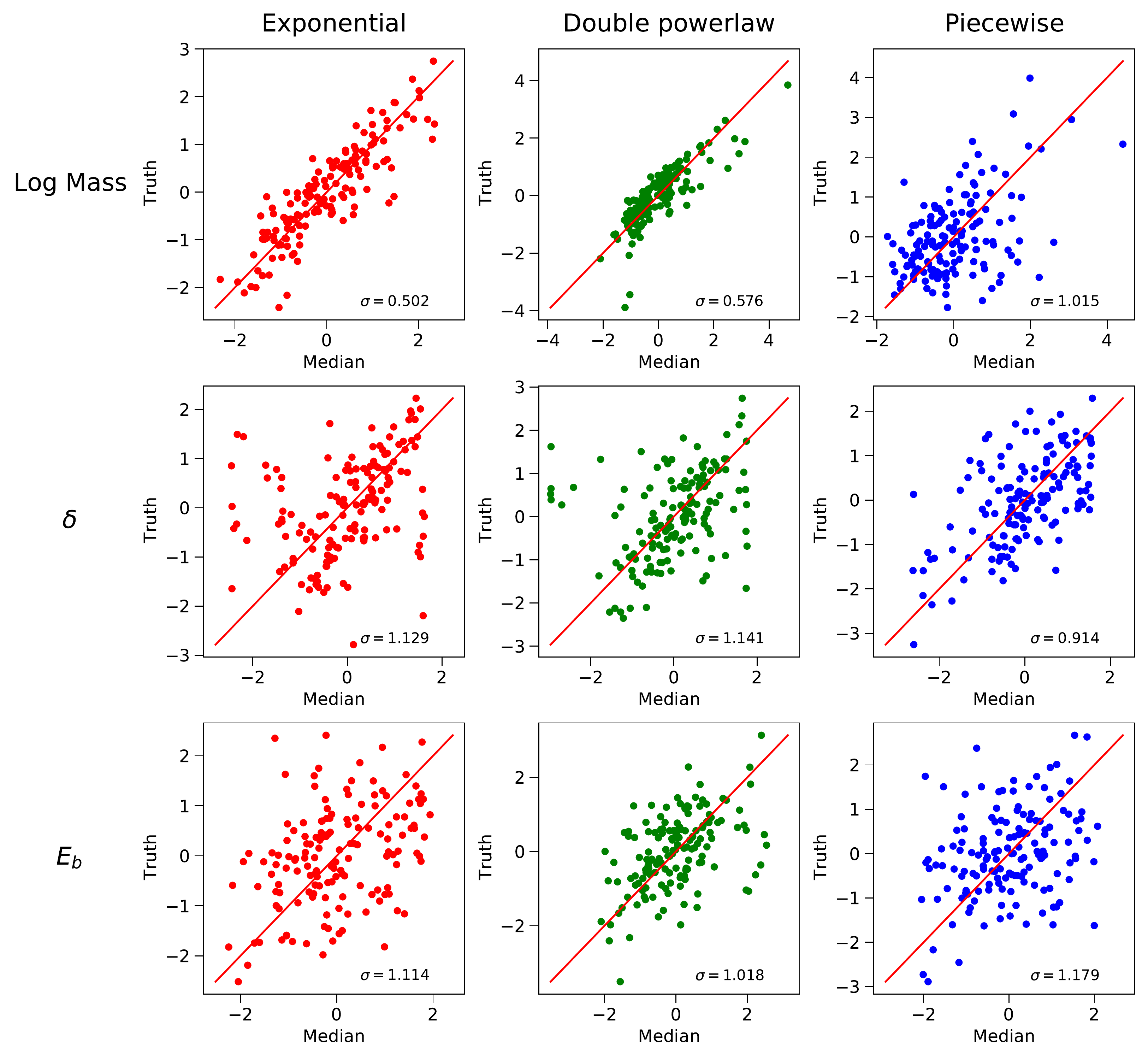}
    \caption{Matrix of plots showing median fitted versus true values for fits to mock galaxies, done without UVOT data. As in Figure~\ref{fig:test_comp}, we are plotting the values standardized by the sample mean and standard deviation. We find confirmation that the additional data increases precision in recovering the dust law parameters. Contrasting with the fits of real data, we do not see a shift in bump strength when removing the UVOT data. The different result seen based on fits to real versus synthetic photometry warrants further discussion and investigation.}
    \label{fig:test_comp_ns}
\end{figure*}

We then fit mock photometry without the UVOT data and compare the resulting parameters to the known values. This mirrors the test we did in Figure~\ref{fig:test_comp} but here we remove UVOT data prior to doing the fitting. Figure~\ref{fig:test_comp_ns} shows the comparison of median fitted values versus true values for fits to mock galaxies without UVOT data. What we find is that the scatter in recovering the true parameters increases as expected. Unlike as seen in Figure~\ref{fig:bump_offset}, where removing UVOT data leads to larger bump strengths for fits to real galaxies, we do not see any bias based on varying the data used for these mock galaxies. This raises questions regarding the deviation seen in the fits of real data and will be discussed more in \S\ref{sec:discussion}.

\section{Discussion}
\label{sec:discussion}

\subsection{Link between SFH and attenuation}

Theoretical work on attenuation laws by \citet{2018ApJ...869...70N} explored the possible attenuation curves arising from a single extinction law and various star-dust geometries. They found that the slope of the attenuation curve was tied to the fraction of unobscured young stars, where more unobscured young stars lead to flatter slopes with weaker bumps. Additionally, a large fraction of unobscured evolved stars leads to a steeper attenuation law. Further, \citet{2018ApJ...869...70N} found that a larger fraction of unobscured O and B stars produced a smaller bump strength. In simpler terms, heavily star forming galaxies are expected to have shallower attenuation curves and a weaker 2175 \AA\ dust bump, as compared to quiescent galaxies, which naturally leads to a relationship between $\delta$ and $E_b$. This points to a physical explanation for the observed relationship between the parameters seen in \citet{Kriek+Conroy} and \citet{2016ApJ...833..201S}. 

The physical relationship between stellar age and dust attenuation implies that the shape of the star formation history will impact the observed dust attenuation law. In SED fitting, the chosen parametric form for the SFH can strongly, and unknowingly, influence the fitted attenuation law due to the implicit ratio of young to old stars imposed by certain SFH shapes. \citet{2018ApJ...869...70N} showed that, in terms of SFH shape, a rising SFH will lead to shallowest laws and a falling SFH will lead to steepest laws. Additionally, varying dust composition will affect these generalizations, making it even more complicated to understand the effects of dust \citep{1977ApJ...217..425M, 2001ApJ...548..296W, Galliano_dustreview}. 

Using this framework, we can attempt to understand the results presented here. We found that while the different SFHs lead to attenuation laws that agreed with each other, the double power law SFH led to a steeper attenuation law and the exponentially declining SFH a more shallow one. The double power law SFH peaks somewhere in the middle of the galaxies' life leading to a moderate stellar age with a significant fraction of unobscured old stars. Combined, one would expect this SFH to lead to steep attenuation laws. The exponential SFH however, would have the oldest mass-weighted age as well as a larger proportion of unobscured old stars. Yet it produced a shallower attenuation curve. In contrast, \citet{2018ApJ...869...70N} found the exponentially declining SFH led to the steepest law. It is possible that due to the asymptotic nature of the decaying exponential SFH, the recent SFH shape is constant rather than declining, leading to a larger fraction of unobscured young stars, which could produce the shallow attenuation law seen.

The piecewise SFH provides the most flexibility, allowing a rising, falling, or constant recent SFH shape. The trade-off of this flexibility is in the number of model parameters, potentially leading to degeneracy between parameters or overfitting the data. Older epochs of star formation are relatively unconstrained by the data, leading to a high uncertainty. However, since we are indirectly inferring the attenuation law in the UV, accurately modeling the most recent epochs of star formation is paramount.

Despite the differences seen, the fits mostly agree with each other regardless of chosen SFH. The discrepancy between the dust curves found here amounts to $\sim0.3$ mag in $A_{1500}$ for $A_V=1$. While we have been careful in our SED fitting and how we analyzed our fits, our sample size could negatively affect our results. However, since the attenuation law's dependence on SFH holds even when possible AGN and galaxies consistent with no dust are removed, we believe the effect is real. Quantifying the effect on a large statistical sample, as well as large sample of similar galaxies, is a subject of future work. 

One shortcoming of our sample is the paucity of galaxies with $\delta<-0.5$ (Figure~\ref{fig:dpl_KC13}). Low-mass or quiescent galaxies are believed to have the steepest attenuation curves \citep{Salim_2018}. Their inclusion would make the sample more robust. \citet{Salim_2018} also finds that local analogs of high redshift galaxies have attenuation laws steeper than the SMC so this region of parameter space becomes even more important to study in the era of {\it JWST}. 

From the mock galaxy fits, we see that the exponential SFH, which is the most simplistic, struggles to be more precise than more complex SFHs. Even when the underlying SFH parameterization matched the one assumed during fitting, the fits involving the exponential SFH had some of the largest scatter between the median and true values. The double power law SFH lead to fits with the best overall precision. However, in real world scenarios, the underlying SFH is likely to be a complicated function. Therefore, we believe that the exponential SFH is too simplistic to model nearby galaxies. More complex SFHs like the double power law or piecewise SFH are likely better able to fit the intricacies of a true SFH. \citet{2019ApJ...873...44C} and \citet{2019ApJ...876....3L} compare parametric and non-parametric SFHs using Prospector and find that non-parametric SFHs, like the piecewise law used here, lead to less bias and more accurate errors for recovering parameters like SFR and stellar mass. These works also highlight the large effect that the chosen priors for each parameter can have on the resulting SED fit. However, overfitting and degeneracies between the large number of parameters can be a problem. While we cannot claim that a single SFH parameterization is the best, care must be used in order to justify the assumptions being made. 

In \citet{lower20}, they found that a delayed exponential star formation history underestimated stellar mass by 0.4 dex while the ``nonparametric" piecewise SFH is much more accurate (scatter of 0.1 dex). This further reinforces the claim that the exponential SFH is not an ideal choice for SED fitting. Although the piecewise SFH adds more parameters to the model, it can lead to more accurate galaxy parameters. The use of more complicated SFHs is becoming more common as evidenced by many studies like \citet{2019ApJ...879..116I, 2019ApJ...877..141M, 2019ApJ...873...44C, 2019ApJ...876....3L}. However, in cases where data are scarce and there are only a few bands of photometry, using a more simplistic model will be necessary to get any meaningful constraints.

Ultimately, the three different attenuation laws here are steeper than the starburst law from \citet{C00}, implying that assuming the starburst attenuation curve would lead to underestimating the intrinsic UV flux of a galaxy based on the results here.

\subsection{Utility of UVOT photometry}

{\it Swift}/UVOT near-UV imaging, including the UVM2 filter centered on the 2175 \AA\ bump, provide a unique opportunity to constrain the dust attenuation law in nearby galaxies. We show a marked improvement in our SED fitting results when including near-UV photometry from {\it Swift}/UVOT. While the slope of the attenuation law can be accurately determined using the broadband coverage from {\it GALEX}, the additional NUV photometry from UVOT tightens constraints on the UV bump. \citet{Decleir2019} found similar utility when including UVOT data, finding that integrated light SED fits without UVOT data led to bump strengths with relative uncertainties above 100\%. 

In addition to UVOT leading to increased precision on parameter estimates, we also observe a shift towards higher bump strengths when excluding the UVOT photometry in fits of our sample. This systematic shift cannot be easily explained. If we were poorly modelling nebular emission, any feature that could systematically affect the bump strength would be captured by both the {\it GALEX} NUV and {\it Swift}/UVOT UVM2 filters. Other possibilities include a UVOT calibration issue such as coincidence loss. However, the shift to higher bump strengths is seen across the sample, regardless of the source brightness.

Interestingly, the shift is not seen when fitting mock galaxies. There are a few possible explanations for this. One is that the higher bump strengths seen without UVOT data indicate that the likelihood is prior-driven and that the data cannot provide reasonable constraints without the additional UVOT data. The shift towards lower bump strengths when including UVOT data would mean that SED fits of this sample rule out solutions with strong, MW-like bumps, and imply that the sample generally exhibits weaker 2175 \AA\ excesses than the MW. Another explanation is that the form of our assumed dust law is incorrect leading to this peculiar result in fits of real data but not in our fits of mock galaxies where the form of the dust law matches perfectly. Lastly, as AGN emission is not modeled, this could have an effect as AGN emission may be present in the fits to real data but is not modeled for our synthetic photometry.

We are inclined to believe that the fits including UVOT data are more accurate due to them better fitting the trend seen in \citet{Kriek+Conroy} between the slope and bump strength. We find that the fits with only {\it GALEX} data lie above this relation and lead to larger bump strengths. This larger bump strength also lines up with the median of our wide, uniform prior, strengthening the argument that the dust law from fits without UVOT data are prior dominated. The trend of larger bump strengths without UVOT data was seen for all three SFHs, further demonstrating the effectiveness of UVOT data in determining the UV bump. The derived bump strengths for the fits with only {\it GALEX} data match the median of the prior meaning that the data cannot constrain the bump strength. A more scientifically motivated prior may yield better results such as having the prior for the bump depend on $\delta$. Lastly, since we see a shift towards lower bump strengths when additional information is included, we conclude that the bump strength is sub-MW for the majority of the sample. 

A likely explanation as to why the {\it GALEX} only fits led to larger, prior-dominated dust bumps is the UVOT data gave a better measurement of the UV continuum near the bump. The three, tightly spaced UVOT filters are situated very closely to 2175 \AA, and this increased sampling of the UV leads to a better fit. As mentioned in \S\ref{sec:SED_fitting}, the two-parameter modified starburst law, while flexible, struggles to accurately reproduce some of the well-measured extinction laws like the SMC law. Using the \citet{C00} law as a base means that the relative difference between the UV and optical slope of the attenuation law is fixed. For the SMC law, matching the optical slope requires a different $\delta$ than for matching the UV slope. A possible fitting quirk could lead to an inflated bump strength to provide a better fit in absence of more UV data. Another possible source of error is our assumption of a fixed location and width of the dust bump. It has been observed that the position and width of the bump can vary \citep{2007ApJ...663..320F, 2022MNRAS.514.1886S}.

Different formulations of the dust law could be explored further. \citet{2022ApJ...931...14L} adds an additional parameter to the modified starburst law that describes the fraction of the stellar light affected by attenuation. This is similar to additional birth cloud reddening for young stellar populations, included here \citep{CF00}. By adding an additional variable for the fraction of unobscured emission, one can mimic more complex star-dust geometries in SED fitting and breaks the uniform screen assumed by \citet{C00}. One issue that may limit the utility of this alternative dust attenuation law formulation is degeneracy between the fraction of unobscured stellar light and the slope $\delta$.

\section{Conclusions} \label{sec:conclusion}

In this work, we present {\it Swift}/UVOT photometry for the SINGS/KINGFISH sample of nearby galaxies. Using these new data and archival data from \citet{Dale2017}, we perform SED fitting to study the diversity of dust attenuation curves and the effect the assumed SFH parameterization has on the derived dust curve. Since the attenuation law can only be inferred indirectly and the SFH is only weakly constrained by the data, we examined how those fitting assumptions affect the results. To do so, we used \verb|MCSED| \citep{2020ApJ...899....7B} to fit the photometry using the modified starburst law from \citet{Noll_09} and three different SFH parameterizations: a declining exponential, a double power law, and a piecewise discontinuous function from \citet{2019ApJ...876....3L}.

Our main results are as follows:
\begin{enumerate}
    \item The average attenuation law from our fits is steeper than the \citet{C00} law for each of the SFHs tested. Additionally, they all agreed with each other withing errors. However, we find some evidence that the SFH affects the derived dust law. Specifically, the exponential SFH preferentially corresponded to the shallowest attenuation curve. The discrepancy between the \citet{C00} law and the attenuation laws derived here suggests a steeper attenuation law with a moderate bump is more appropriate to use when modeling attenuation of galaxies in the local universe.
    \item We generate and fit mock photometry and find that the fits using the exponential SFH have the lowest precision in recovering galaxy properties like stellar mass and the dust law parameters and struggles to outperform the more complex piecewise SFH. Based on other results in the literature and the performance seen here, more complex SFHs like the piecewise or double power law SFH should be the standard in SED modeling \citep{2019ApJ...873...44C, 2019ApJ...876....3L, lower20}.
    \item When comparing SED fits with and without UVOT photometry, we find that the inclusion of the UVOT data increases the precision on 2175 \AA\ bump strength measurements and also, on average, leads to lower bump strength estimates, which are more in line with literature trends \citep{Kriek+Conroy}. The spread of derived bump strengths from {\it GALEX}-only fits matches the prior, implying that obtaining meaningful constraints on the bump is challenging without better informed priors or more data like UVOT photometry. Therefore, UVOT data are crucial to be able to model the UV bump in nearby galaxies via SED fitting.
    \item Lastly, the discrepancy where fits including UVOT lead to systematically lower bump strengths than fits without does not appear when fitting mock data. While we believe the shift seen is since the bump strength estimate is prior-dominated, another possible explanation could be an issue in the assumptions made. For example, AGN emission could be a factor or our chosen attenuation law parameterization could be problematic. While the \citet{Noll_09} law is flexible, its reliance on the \citet{C00} law causes it to struggle to capture all the observed intricacies of the dust curves of the Magellanic clouds. Other formulations of the attenuation law like the one proposed in \citet{2022ApJ...931...14L} could be explored in future work. 
\end{enumerate}

One of the main limitations of this work is the sample size; using a much larger sample of galaxies with UVOT data would improve the statistical power of this analysis. For example, the Local Volume Legacy sample is a volume-limited sample of nearby galaxies that will be a good testing ground for a larger statistical study of the effects of SFH, demonstrating the utility of UVOT data, and understanding population level trends regarding dust attenuation \citep{LVL}. The SINGS/KINGFISH sample lacks enough galaxies with very steep attenuation laws, corresponding to low-mass quiescent galaxies. Controlling for the selection effects is not possible with our current sample so a larger number of galaxies are needed. 

In a future work, Decleir et al. (in prep.), we will compare resulting SED fits across different codes like \texttt{MCSED}, CIGALE, and Prospector. Additionally, we will look for correlations with other galaxy properties as seen in the literature \citep{Kriek+Conroy, 2017ApJ...851...90B, Salim_2018, decleir_thesis, 2020ApJ...899..117S}.  

Like other works studying dust in nearby galaxies, SED fitting on a pixel by pixel basis would aid in the understanding of the spatial distribution and geometry of dust as well as the effects of galaxy environment. \citet{2021ApJS..254...15A} introduced piXedfit, a code to combine different resolution photometry as well as integral field spectroscopy data for SED fitting and is being used in resolved studies \citep{2022ApJ...926...81A, 2022ApJ...935...98A}. Many studies like \citet{Decleir2019} have used pixel-by-pixel SED fitting to explore the effects of galaxy environment and some have used IFU spectroscopy like \citet{mallory} to understand the relationship between stellar and nebular attenuation. Lastly, since we have shown that {\it GALEX} photometry alone cannot constrain the bump beyond the prior, more sophisticated SED fitting codes like Prospector could be used as it would allow more informative priors to be used for the dust law parameters, in particular $E_b$ \citep{2021ApJS..254...22J}.

Here, we have shown that there is a slight dependence on assumed SFH when deriving the attenuation law via SED fitting. However, when comparing the steepest derived law here to the starburst law, the difference in $A_{1500}$ for $A_V=1$ is greater than a magnitude. This can lead to significant errors that greatly affect derived galaxy properties. Correcting for dust attenuation in the UV is extremely important when studying high redshift galaxies with telescopes like {\it JWST}, where longer wavelength emission not affected by dust attenuation is not easily observable. In these cases, it is imperative to have an accurate dust correction to do the highest precision science possible.

\acknowledgments{
The authors thank the anonymous referee for their
comments and suggestions which improved the clarity of this work. We gratefully acknowledge the efforts of Erik Hoversten to acquire the {\it Swift}/UVOT data of the SINGS/KINGFISH galaxies and begin this project. We also thank Robin Ciardullo, Gautam Nagaraj, Mallory Molina, Joel Leja, Hyungsuk Tak and Danny Dale for many helpful conversations and suggestions that greatly improved this paper. 

The authors acknowledge sponsorship at PSU by NASA contract NAS5-00136. We also acknowledge funding through NASA ADAP 15-ADAP15-0131.  

This research has made use of NASA's Astrophysics Data System. This research has made use of data and/or software provided by the High Energy Astrophysics Science Archive Research Center (HEASARC), which is a service of the Astrophysics Science Division at NASA/GSFC and the High Energy Astrophysics Division of the Smithsonian Astrophysical Observatory.
\software{Astropy \citep{2018AJ....156..123A, 2013A&A...558A..33A}, matplotlib \citep{Hunter:2007},  NumPy \citep{harris2020array}, pandas \citep{McKinney_2010}, and SciPy \citep{Virtanen_2020}.} }

\bibliography{sample63}{}
\bibliographystyle{aasjournal}

\end{document}